\documentclass[pdflatex,sn-mathphys-num]{sn-jnl}

\usepackage{graphicx}%
\usepackage{multirow}%
\usepackage{amsmath,amssymb,amsfonts}%
\usepackage{amsthm}%
\usepackage{mathrsfs}%
\usepackage[title]{appendix}%
\usepackage{xcolor}%
\usepackage{textcomp}%
\usepackage{manyfoot}%
\usepackage{booktabs}%
\usepackage{algorithm}%
\usepackage{algorithmicx}%
\usepackage{algpseudocode}%
\usepackage{listings}%
\usepackage{cleveref}
\usepackage{subcaption}
\usepackage{siunitx}
\usepackage{tikz}

\theoremstyle{thmstyleone}%
\theoremstyle{thmstyletwo}%

\theoremstyle{thmstylethree}%

\begin{document}

\title[Article Title]{Detecting Turbulent Patterns in Particulate Pipe Flow by Streak Angle Visualization}

\author[1]{\fnm{Rishav} \sur{Raj}}\email{rishavrajrnc@gmail.com}

\author[1]{\fnm{Abhiram} \sur{Thiruthummal}}\email{thiruthuma@uni.coventry.ac.uk}

\author*[1]{\fnm{Alban} \sur{Poth\'erat}}\email{alban.potherat@coventry.ac.uk}

\affil*[1]{\orgdiv{Centre of Fluids and Complex Systems}, \orgname{Coventry University}, \orgaddress{\street{Maudslay House, Mile Lane}, \city{Coventry}, \postcode{CV1 2NL}, \country{United Kingdom}}}

\abstract{Detecting the transition from laminar to turbulent flow in particulate pipe systems remains a complex issue in fluid dynamics, often requiring sophisticated and costly experimental apparatus. This research presents an innovative streak visualization method designed to offer a simple and robust approach to identify transitional turbulent patterns in particulate pipe flows with neutrally buoyant particles. The technique employs a laser arrangement and a low-cost camera setup to capture particle-generated streaks within the fluid, enabling real-time observation of flow patterns. Validation of the proposed method was conducted through comparison with established techniques like Particle Image Velocimetry (PIV) and pressure drop measurements, confirming its accuracy and reliability. Experiments demonstrate the streak visualization method's capacity to differentiate between laminar, transitional, and turbulent flow regimes by analyzing the standard deviation of streak angles. The method is especially efficient at low particle concentration, ie precisely where other more established methods become less effective. Furthermore, this technique enables us to identify a critical Reynolds number using Kullback-Leibler divergence built on the statistical distribution of streak angles, which is consistent with previous studies.
Because of it is effective at low concentrations and robust, this streak visualization technique opens new perspectives for the characterization of particulate pipe flows not only in the confines of the laboratory, but also in less controlled industrial multi-phase flows where determining the laminar or turbulent nature of the flow is a prerequisite for flowmeter calibration. 
}




\maketitle

\section{Introduction }

We introduce a particle streak-based visualization technique that is simple, cost-effective, and accurate for studying turbulence patterns and flow transitions in particulate pipe flows. The idea is to use minimal instrumentation detecting particles only to circumvent the need for separation between particles and fluid phase and offer a simple, robust method that is easy to implement in a wide variety of environments. Hence, 
This method provides an accessible solution for research in engineering applications such as industrial pipelines, biological systems, and environmental processes \citep{breithaupt2021visualization, pozniak2017business, post2002feature, bujack2020state}.
By analyzing particle streak angles, which reflect trajectory behavior \citep{Zhou:16, kumar2020applying}, we use statistical tools to characterize flow states and detect turbulence with minimal disruption.

Classical single phase pipe flows dynamics exhibit subcritical transitions, where turbulence arises abruptly from finite-amplitude perturbations 
\citep{salwen1972stability, salwen1980linear}. While infinitesimal disturbances decay due to stability at all Reynolds numbers, finite amplitude perturbations of sufficient amplitude trigger turbulence through nonlinear mechanisms. This mechanism allows laminar and turbulent states to coexist \citep{hof2003scaling, manneville2015subcritical} so the transition to turbulence involves localized turbulent structures such as "puffs" and "slugs," which propagate through the pipe \citep{wygnanski1973transition, mullin2006recent, eckhardt2007turbulence}. Puffs remain confined in transitional regimes, while slugs expand and lead to fully developed turbulence at higher Reynolds numbers \citep{barkley2016rise, avila2011onset, barkley2016theoretical}.

\begin{table}[h!]
\centering
\renewcommand{\arraystretch}{1.3} 
\setlength{\tabcolsep}{3pt} 
\resizebox{\textwidth}{!}{ 
\begin{normalsize} 
\begin{tabular}{|p{5cm}|p{6cm}|p{6cm}|}
\hline
\textbf{Technique}                & \textbf{Measured Parameters (Two-Phase Flow)}                & \textbf{Limitations} \\ \hline
Dye Injection                     & Qualitative visualization of flow patterns and mixing regions. & No velocity quantification; ineffective in uniform-density flows. \\ \hline
Schlieren Imaging                 & Visualizes density gradients and large-scale flow structures. & Ineffective in flows without significant refractive index or density variations; qualitative only. \\ \hline
Hot-Wire Anemometry               & Fluid Velocity using temperature fluctuations     & Intrusive; unsuitable for particle-laden flows; cannot differentiate between particle and fluid velocities. \\ \hline
Particle Image Velocimetry (PIV)  & Velocity fields of fluid phase using seeded particles.       & High cost; computationally intensive; limited spatial coverage; particles must follow flow accurately. \\ \hline
Laser Doppler Velocimetry (LDV)   & Point-wise velocity measurements of fluid phase.             & Limited to single points; requires precise alignment; costly; unsuitable for dense particle suspensions. \\ \hline
Planar Laser-Induced Fluorescence (PLIF) & Concentration and scalar fields such as density and temperature. & Ineffective for direct velocity data; fluorescence quenching; expensive and calibration-intensive. \\ \hline
Ultrasound Image Velocimetry (UIV) & Velocity distribution of fluid phase in opaque flows.        & Limited spatial resolution; calibration difficulties in non-homogeneous particle-fluid systems. \\ \hline
Magnetic Resonance Velocimetry (MRV) & 3D velocity profiles of fluid phase; non-invasive.          & Very slow; high cost; unsuitable for transient phenomena or rapid flow transitions. \\ \hline
\end{tabular}
\end{normalsize} 
}
\caption{Comparison of Flow Visualization Techniques Their Limitations in Particulate Pipe Flow Systems}
\label{table1}
\end{table}

This classical problem of the laminar-turbulent transition in single-phase flows is critical for predicting fluid behavior, with Reynolds number as a key factor \citep{schmid2002stability, pope2001turbulent}. Yet, many applications involve the presence of particles which may alter this scenario. 
Indeed, introducing even neutrally buoyant particles add complexity by modifying flow stability, altering thresholds for transition, and influencing the behavior of turbulent structures \citep{matas2003transition, agrawal2019transition}. At low concentration, neutrally buoyant particles tend to migrate to specific radial locations\cite{segre1962_jfm,matas2004_pf}, under the effect of the lift force incurred by shear-induced rotation \cite{oliver1962_nat,repetti1964_nat,schonberg1989_jfm,hogg1994_jfm,han1999_jr,asmolov1999_jsme}. This so-called \emph{Segr\'e-Silberberg} effect may even cause the base parabolic Poiseuille velocity profile to become linearly unstable at Reynolds numbers as low as 400 \cite{rouquier2019_jfm}, but the subsequent nonlinear development of this instability is still unclear.
In practice, these effects are important for applications ranging from pipelines to environmental systems, where particle-fluid interactions play a significant role, and where knowledge of the flow state is a key requirement to calibrate flowmeters \citep{hogendoorn2022onset, leskovec2020pipe}.

Traditional techniques for studying the stability of particulate pipe flow experimentally face significant limitations. Dye injection and schlieren imaging offer qualitative observations but lack quantitative velocity data and are ineffective for detecting fine-scale features, such as the Segre-Silberg effect  \citep{post1991fluid, settles1986modern, settles2001schlieren, merzkirch1987flow, tropea2007springer}. Hot-wire anemometry provides high temporal resolution but is unsuitable for particle-laden flows due to contamination risks and the inability to differentiate particle and fluid velocities \citep{comte1976hotwire, bradshaw1971turbulence}. Its intrusive nature also disturbs the flow, limiting its effectiveness for sensitive transitional phenomena where finite amplitude perturbations need to be controlled \citep{ferrari2022review}.

Modern techniques such as Particle Image Velocimetry (PIV) and Laser Doppler Velocimetry (LDV) are widely used for quantitative measurements. PIV, a planar optical technique, offers high-resolution velocity fields by analyzing the motion of seeded particles within the flow \citep{adrian1991particle, adrian2005twenty}. However, it requires costly and complex setups involving high-power lasers, relatively high-speed cameras, and synchronization systems, which limit its applicability outside dedicated research laboratories \citep{raffel2018particle}. Additionally, the data processing is computationally demanding, and the system is sensitive to misalignments and calibration errors. LDV provides point-wise velocity measurements with high temporal resolution but is limited to single-point data acquisition, making it unsuitable for analyzing large-scale or spatially distributed flow features \citep{george1973laser, albrecht2013laser}. Like PIV, LDV relies on high-power lasers and requires meticulous optical alignment, further complicating its use outside the lab\citep{adrian2017laser, coupland2000laser}.

Advanced techniques like Planar Laser-Induced Fluorescence (PLIF), Ultrasound Image Velocimetry (UIV), and Magnetic Resonance Velocimetry (MRV) are widely employed to study particle-laden flows, each with specific advantages and limitations. PLIF measures scalar fields such as concentration or temperature using laser-induced fluorescence but is unsuitable for direct velocity measurements and faces challenges like fluorescence quenching, complex calibration, and high operational costs \citep{schultz1972laser, crimaldi2008planar}. UIV tracks acoustic scatterers using ultrasound waves, making it particularly suitable for optically opaque or dense particle-laden systems, though it is limited by spatial resolution and calibration difficulties in non-homogeneous flows \citep{dash2022ultrasound, poelma2017ultrasound}. This technique however fails for diluted particles concentrations, especially where localisation may leave large regions of the flow free of particles. MRV offers 3D velocity measurements without optical access, making it ideal for non-transparent systems, but it is hindered by high costs and long acquisition times \citep{elkins2007magnetic, hartung2011magnetic}. While these methods are effective, their reliance on expensive and complex setups restricts their applicability. Furthermore, while the extensive data they deliver makes them suitable for detailed mapping of the flow, the limited purpose of deteting flow patterns may be fulfilled with much less extensive datasets delivered by simpler setups.
Additionally, existing methods often struggle to capture transient structures like puffs and slugs over the full pipe length due to high costs, accessibility issues, or safety constraints. As shown in Table~\ref{table1}, various flow visualization techniques have distinct limitations in particulate pipe flow systems.
These challenges stress the need for scalable, versatile solutions for both research and industrial applications.

Most existing methods are impeded by the coexistence of the solid phase (particles) and the liquid phase. To circumvent these issues, we propose to use a less sensitive setup that only detects particles, and that is not affected by the fluid phase. The main idea is that where turbulence exists in the fluid phase, neutrally-buoyant particles should follow erratic trajectories, whereas in laminar regions their trajectories should be close to straight lines aligned with the pipe axis. The challenges here are to introduce a meaningful way to mathematically distinguish these two types of trajectories, and also to verify that they reliably map to turbulent and laminar states of the fluid phase, respectively.
For these purposes, the streak-based visualization technique we present in this paper combines simplicity with robust statistical tools. Using the standard deviation of particle streak angles, we classify flow regimes and turbulent features, while the Kullback-Leibler divergence provides a novel method for determining the critical Reynolds number, effectively capturing the laminar-to-turbulent transition. This approach is particularly suited for low-particle-concentration flows and transparent fluids, where traditional methods are less effective. By addressing these current limitations and offering a cost-effective alternative, this method offers new perspectives to study transitional particulate flows at low concentration. Its principle may also be implemented to develop instrumentation capable of rapdidly identifying turbulent pattern in constrained industrial environment.

The paper is laid out as follows: we first describe the experimental setup (section \ref{sec2}, the streak-visualisation system and associated data processing technique (section \ref{sec3}). We then validate the identification of flow patterns against classical PIV, PTV and pressure measurement techniques. Finally, we show how a simple measure of the standard deviation of the angle of particle trajectories enables us to detect turbulence flow patterns. We also show that more refined properties of the statistical distribution of these angles offer a way to define a critical Reynolds for the transition to turbulence consistent with previous studies (section \ref{streak}).

\section{Experimental Setup}\label{sec2}
\begin{figure}[h]
  \begin{center}
    \includegraphics[width=\textwidth]{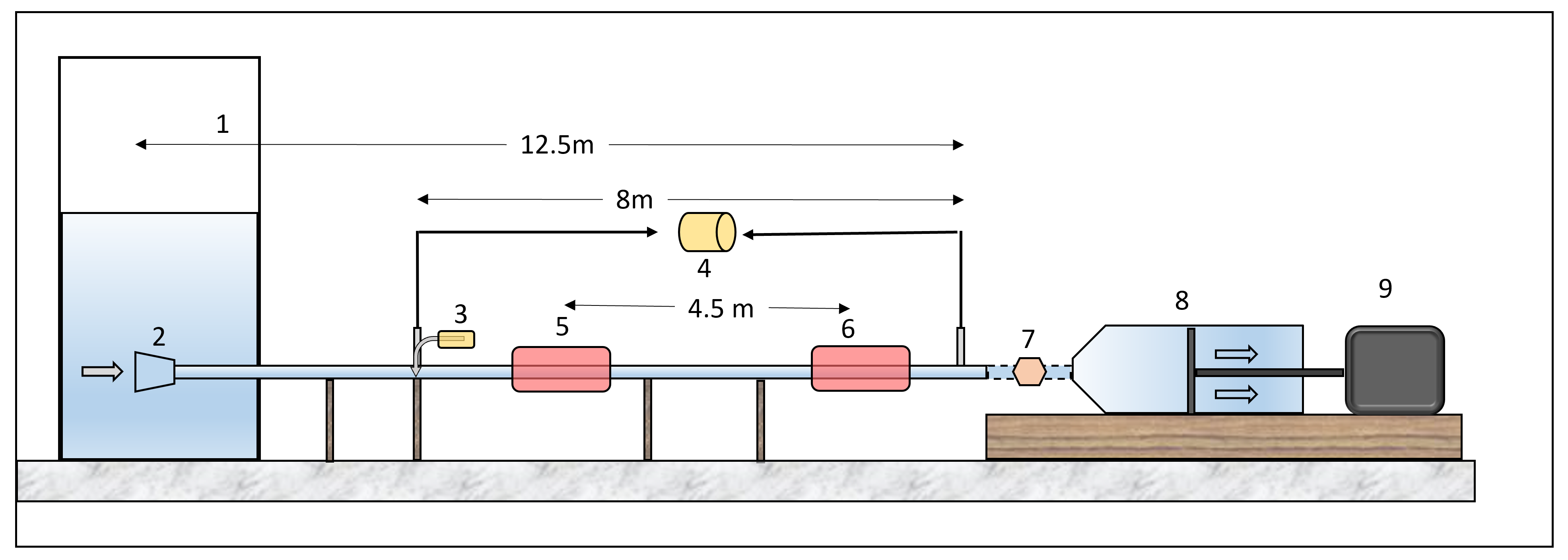}
  \end{center}
  \caption{A 2-D schematic diagram of the rig. (1) Fluid Reservoir, (2) Bell-mouth inlet,(3) Perturbation system, (4) Differential pressure meter, (5) $1^{st}$ visualization system, (6) $2^{nd}$ visualization system, (7) Mass flow meter, (8) Piston-cylinder arrangement, (9) Motor}
  \label{setup}
\end{figure}

The experimental setup {is an upgrade} of the setup described in detail in \citep{singh2020simultaneous}. It {consists} of several subsystems, all represented in \Cref{setup}.
The pipe and hydraulic elements {form} the main component, through which the fluid and the particles {travel}. The glass pipe assembly {comprises} 10 cylindrical borosilicate glass tubes, each \SI{1.2}{\metre} long, with a bell mouth inlet and an additional \SI{25}{\centi\metre} glass section at the outlet. The pipe sections {are} of lengths \SI{1.2}{\metre} $\pm$ \SIrange{10}{30}{\micro\metre}, an inner bore diameter of $D = \SI{20}{\milli\metre} \, \pm \, \SI{0.01}{\milli\metre}$
, and a wall thickness of \SI{3.1}{\milli\metre} $\pm$ \SI{0.03}{\milli\metre} {and are manufactured} to high precision to minimize disturbances caused by geometric irregularities. This setup {is designed} to experimentally study the effects of solid, neutrally buoyant, spherical particles on the transition to turbulence in dilute particle-laden pipe flows.

The first upgrade to the original system {is} a change of working fluid, tracers, and particles that {makes} the rig now operable with water with the addition of a small amount of glycerol used to precisely match fluid and particle densities. This adjustment {results} in a fluid density (\(\rho_f\)) of \SI{1000}{\kilogram\per\cubic\metre}, composed of \(98.4\%\) water and \(1.6\%\) glycerol. The dynamic viscosity of the final solution {is approximately} $\mu=\SI{1.13e-3}{\pascal\second}$.

This mixture {is} easier to handle than the aqueous solution of Sodium Polytungstate that {was initially chosen} to match the density (\(2500 \, \text{kg/m}^3\)) of glass particles used to seed it. This change {is made} possible by the availability of opaque polyethylene particles of density close to that of water. Two types of particles {are required} to study particulate flows: the first ones {are tiny} silver-coated hollow glass particles (in the size range of \SI{10}{\micro\metre}) that {follow} the flow almost instantaneously and {are used} for Particle Image Velocimetry. For the remainder of this paper, these particles {are referred} to as \emph{tracers}. To act like tracers, the particles' response time {needs to be} much smaller than the time scale of the flow, or equivalently, their Stokes number $St=\rho d_p^2 U/(18 \mu D)$ {has to be} much smaller than unity.

In our experiments, the tracers' Stokes number is within the range \([7 \times 10^{-6}, 2 \times 10^{-4}]\), and so fulfills this criterion. 

The second type of particles, on the other hand, {are} much larger particles, with St \text{ close to } unity, alter the dynamics of the fluid phase, and {are used} to study the particulate flow dynamics that result from the interaction between those two phases. Two ranges of diameter particles are utilized: the first type has diameters $d_p$ ranging from \SIrange{425}{500}{\micro\metre}, while the second type ranges from \SIrange{212}{250}{\micro\metre}. The particle-to-pipe diameter ratios are respectively between \([0.0212, 0.025]\) and \([0.0106, 0.0125]\). The particle concentration during experiments {is kept} at \(C = \SI{1.2}{\kilogram\per\cubic\metre}\), corresponding to a volume fraction of \(\Phi = 1.2 \times 10^{-3}\) in all cases.

Upstream of the pipe, the system {is fed} by a reservoir where the fluid {is stored}. A bell mouth inlet placed inside this reservoir {allows} a smooth entry of the fluid into the pipe. The other end of the pipe {is connected} to a piston-cylinder arrangement driven by a motor. The fluid from the reservoir {is pulled} through the pipe when this motor pulls the piston, thus creating a pressure difference and driving the flow.

The second subsystem {is aimed} at introducing precisely controlled velocity perturbations perpendicular to the mean flow. This perturbation subsystem {is made} of a syringe attached to a stepper motor, connected to the main pipe \SI{4.5}{\metre} downstream of the inlet. The diameter of the perturbation injection inlet {is} $d_{\mathrm{pert}} = \SI{2.2}{\milli\meter}$. The stepper motor {is controlled} by an Arduino that {sets} the volume and flow rate of the injected perturbation. The effect of the introduced perturbation on the fluid-particle system {is visualized} at two downstream locations by two different measurement systems.

The 1\textsuperscript{st} visualization system {consists} of a powerful \SI{1}{\watt} laser and a high-speed camera {used} for simultaneous Particle Tracking Velocimetry (PTV) and Particle Image Velocimetry (PIV). The purpose of this system {is} to independently and simultaneously track large particles and map flow velocities by PIV in a vertical plane aligned with the pipe axis, lit by the laser. The visualization section {is located} 400 pipe diameters (\SI{8}{\metre}) from the center of the section to the pipe inlet. The details of this technique {can be found} in Ref. \citep{singh2020simultaneous}.

The newly developed streak visualization system, positioned as the $2^{\text{nd}}$ visualization setup, {is centered} 225 pipe diameters (\SI{4.5}{\metre}) downstream from the center of the $1^{\text{st}}$ visualization system, following the PTV/PIV setup.

\begin{figure}[h]
  \begin{center}
    \includegraphics[width=\textwidth]{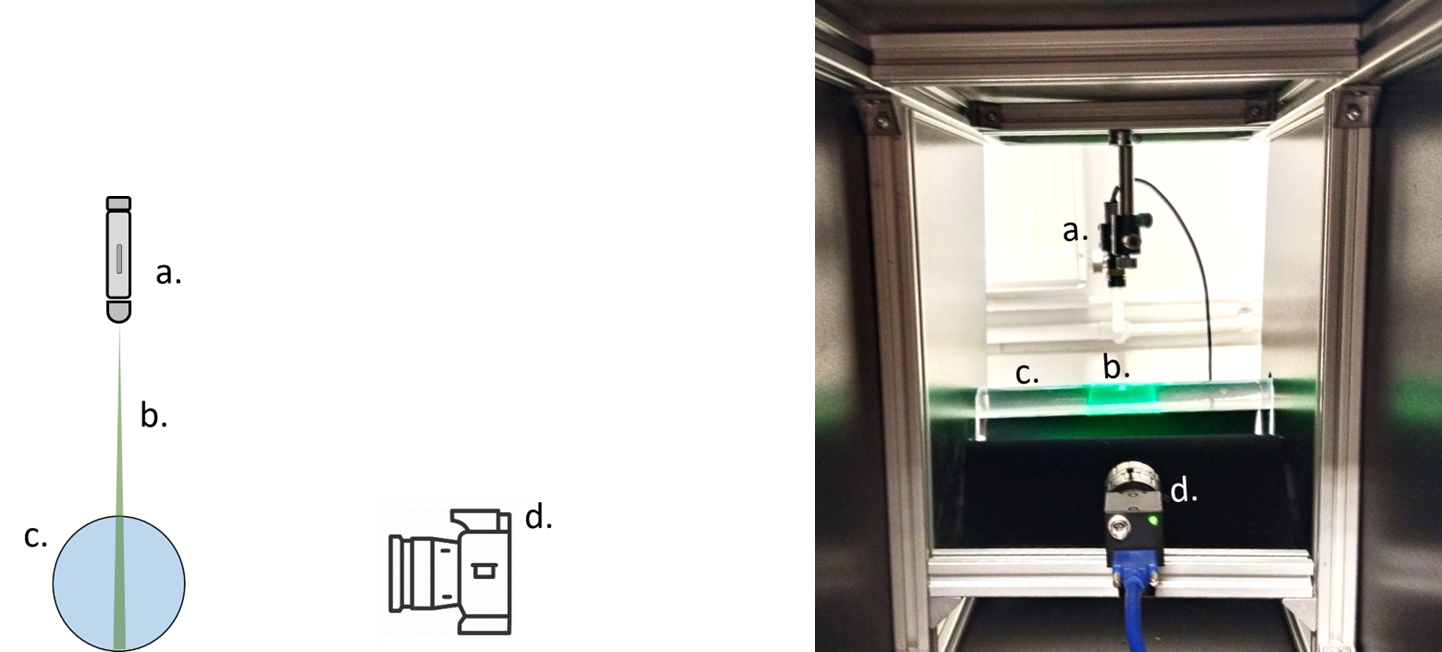}
  \end{center}
  \caption{2-D diagram of the lateral view of streak visualization setup (left) and picture of the actual streak visualization system (right) for streak-angle velocimetry. (a) laser, (b) laser sheet, (c) glass pipe, and (d) camera and lens.}
  \label{fig:side_by_side}
\end{figure}

\Cref{fig:side_by_side} {shows} a 2-D lateral view and a picture of this streak visualization setup placed in the rig. The visualization system {employs} a \SI{50}{\milli\watt} laser and a Flea camera (Blackfly-FLIR). The camera's resolution {is set} to 1536 x 2048 pixels, and it {is equipped} with a \SI{35}{\milli\metre} focal length lens. The sensor size {is approximately} \(4.8 \times 10^{-3} \, \text{m} \times 3.6 \times 10^{-3} \, \text{m}\). The gain {is adjusted} to prevent excessive brightness. The opaque particles, introduced into the fluid under examination, {reflect} the laser light sheet. The camera {is oriented} perpendicular to the laser sheet and {captures} images with adjusted exposure time varying from \SIrange{15}{50}{\milli\second}, for the range of Reynolds number considered in this paper \( Re = \frac{UD}{\nu} \in [1120, 2980] \). Here \( U \), used to calculate the Reynolds number, {is} the average velocity of the fluid across the entire cross-section of the pipe, {calculated} by dividing the volumetric flow rate by the cross-sectional area.

This camera setting {is chosen} to ensure that the captured images depict the particles as distinct streaks of light, thus providing a visual representation of the angles subtended by the path of the particles, with respect to the pipe axis.

Lastly, the test rig {is also fitted} with a high-speed differential pressure transducer (Omega USBH Series). It {is} a USB-based device and {comes} with its own software for direct recording in the computer. It {has} a range of \SIrange{0}{70}{\milli\bar}, with an accuracy of \(\pm 0.08\%\) and a sampling frequency of \SI{1000}{\hertz}.
The differential pressure {is measured} across the pipe between the point of injection of the disturbance and the pipe’s end section over a distance of \SI{8.3}{\metre}. An extra \SI{20}{\centi\metre} glass section with a T-inlet {is manufactured} separately and {connected} at the end of the pipe to connect it there.

\begin{figure}[H]
    \centering
    \begin{subfigure}{0.45\textwidth}
        \includegraphics[width=\linewidth]{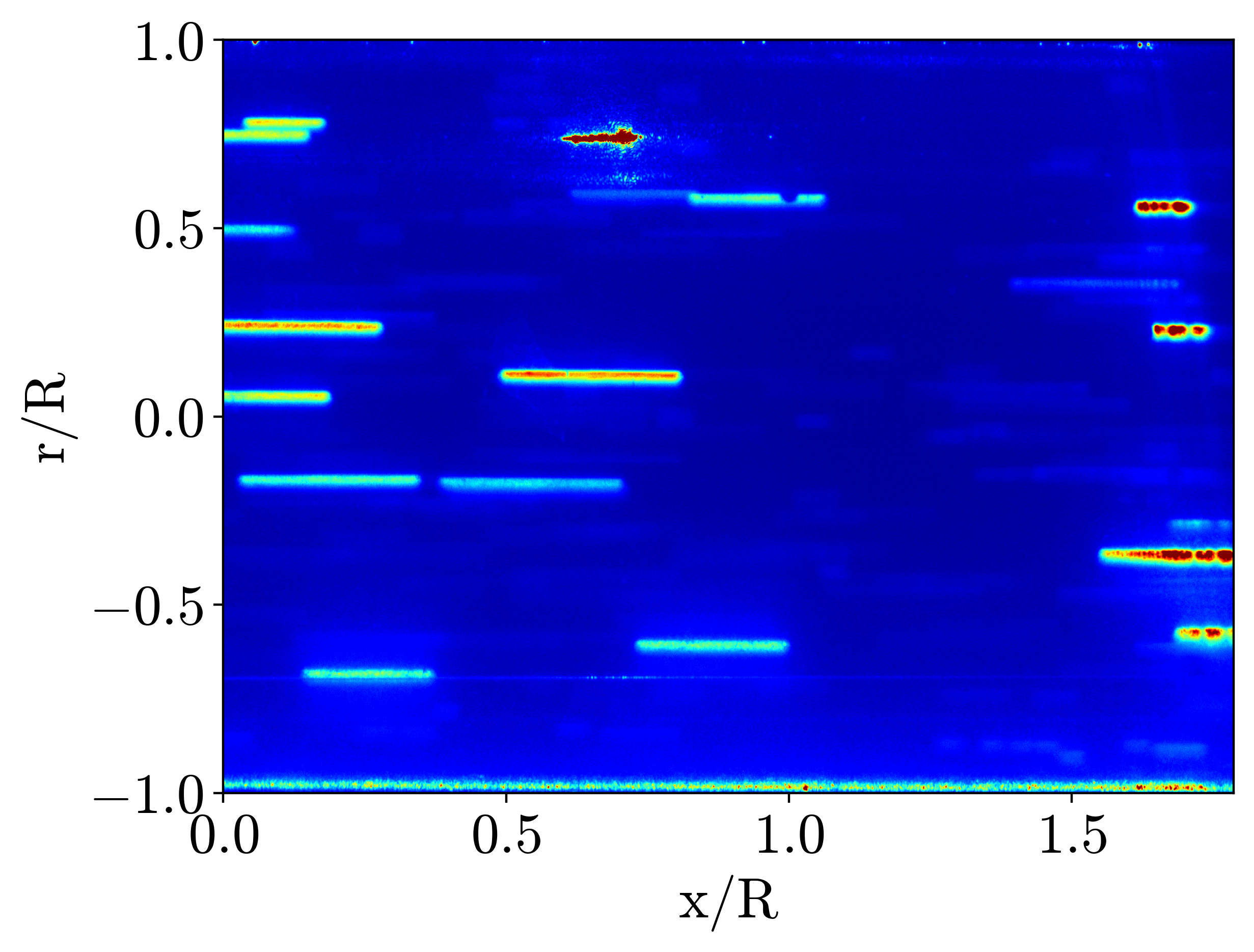}
        \caption{}
        \label{fig:raw_image}
    \end{subfigure}
    \begin{subfigure}{0.45\textwidth}
        \includegraphics[width=\linewidth]{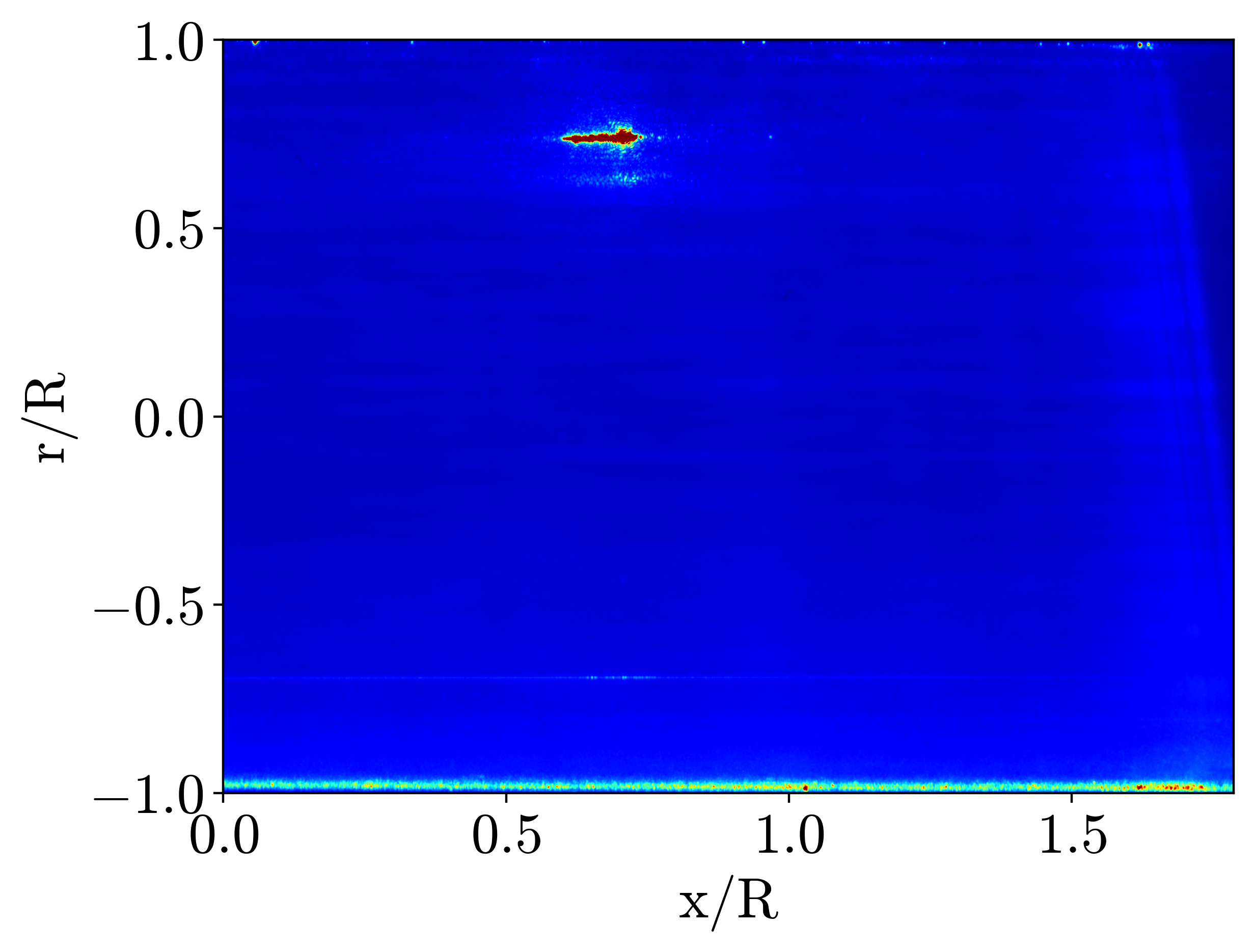}
        \caption{}
        \label{fig:background_image}
    \end{subfigure}
    \begin{subfigure}{0.45\textwidth}
        \includegraphics[width=\linewidth]{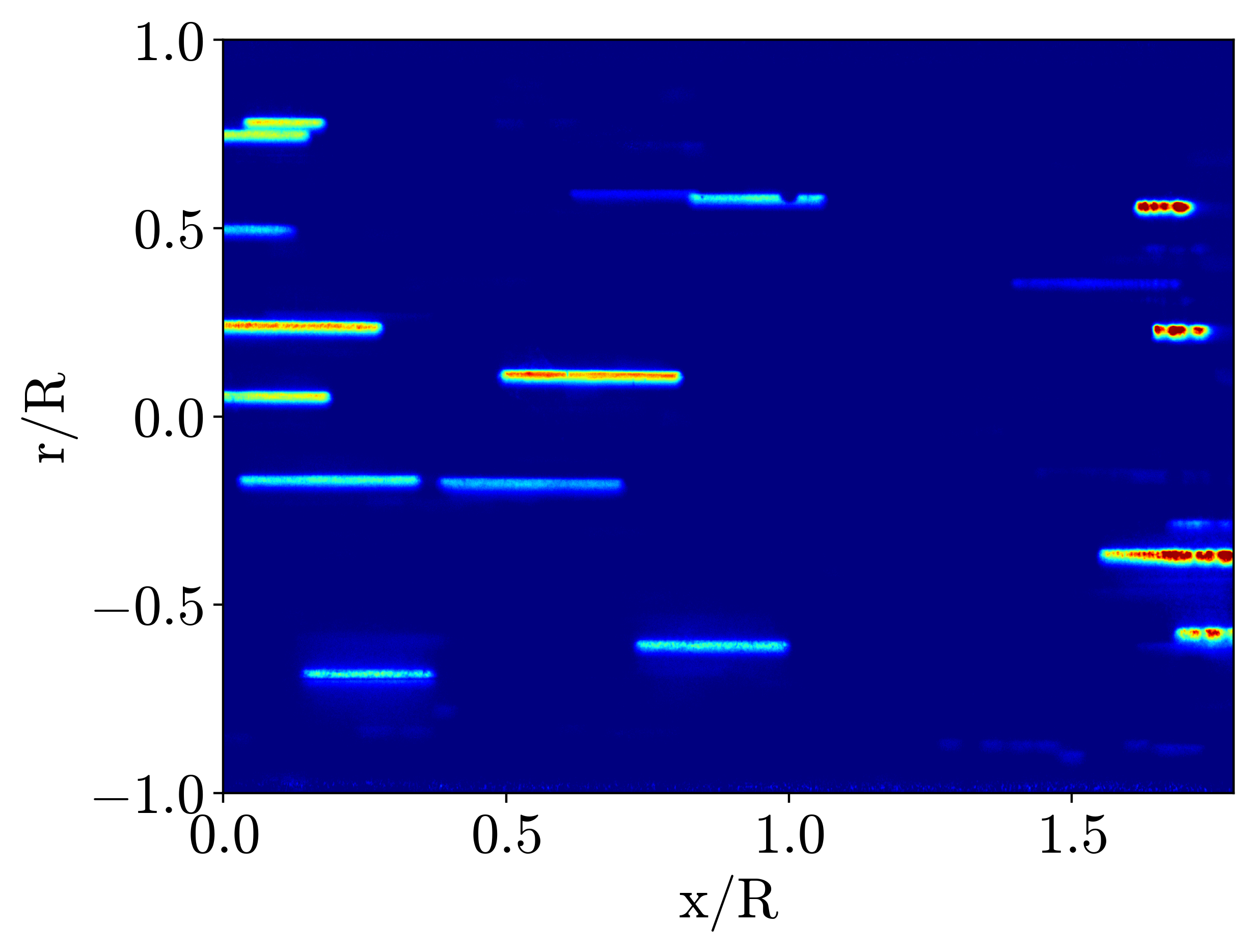}
        \caption{}
        \label{fig:background_subtracted_image}
    \end{subfigure}
    \begin{subfigure}{0.45\textwidth}
        \includegraphics[width=\linewidth]{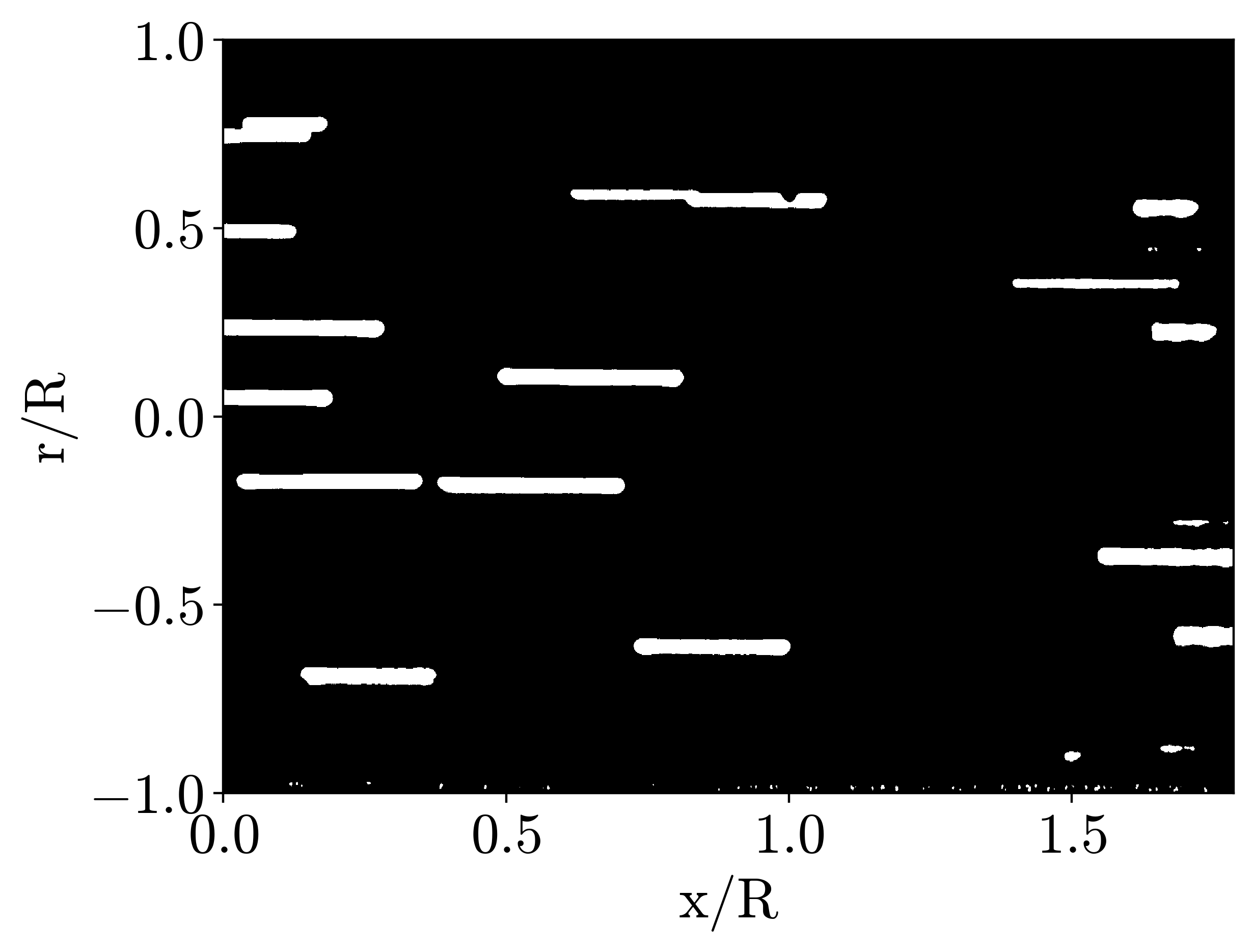}
        \caption{}
        \label{fig:adaptive_threshold}
    \end{subfigure}
    \begin{subfigure}{0.45\textwidth}
        \includegraphics[width=\linewidth]{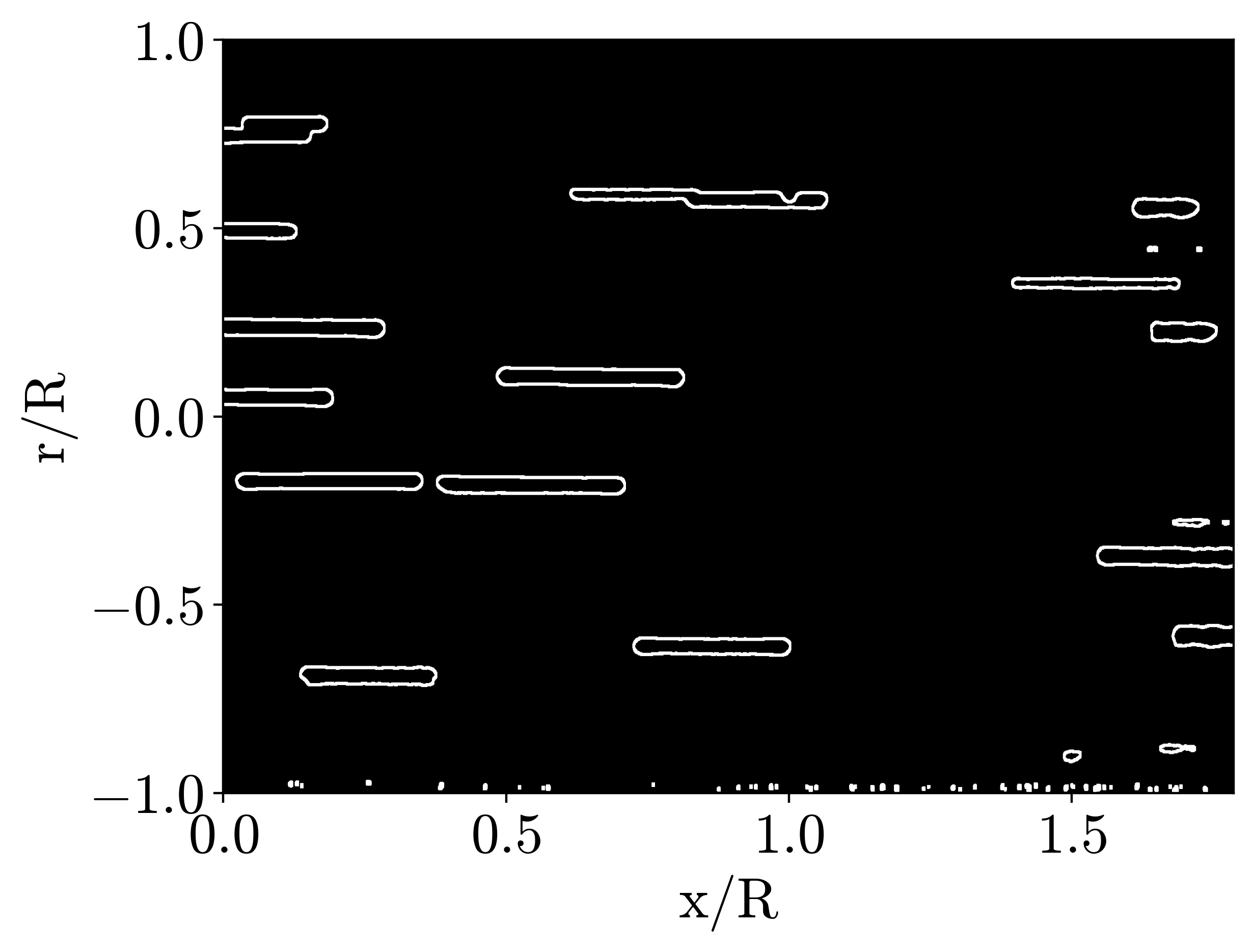}
        \caption{}
        \label{fig:edge_detection}
    \end{subfigure}
    \begin{subfigure}{0.45\textwidth}
        \includegraphics[width=\linewidth]{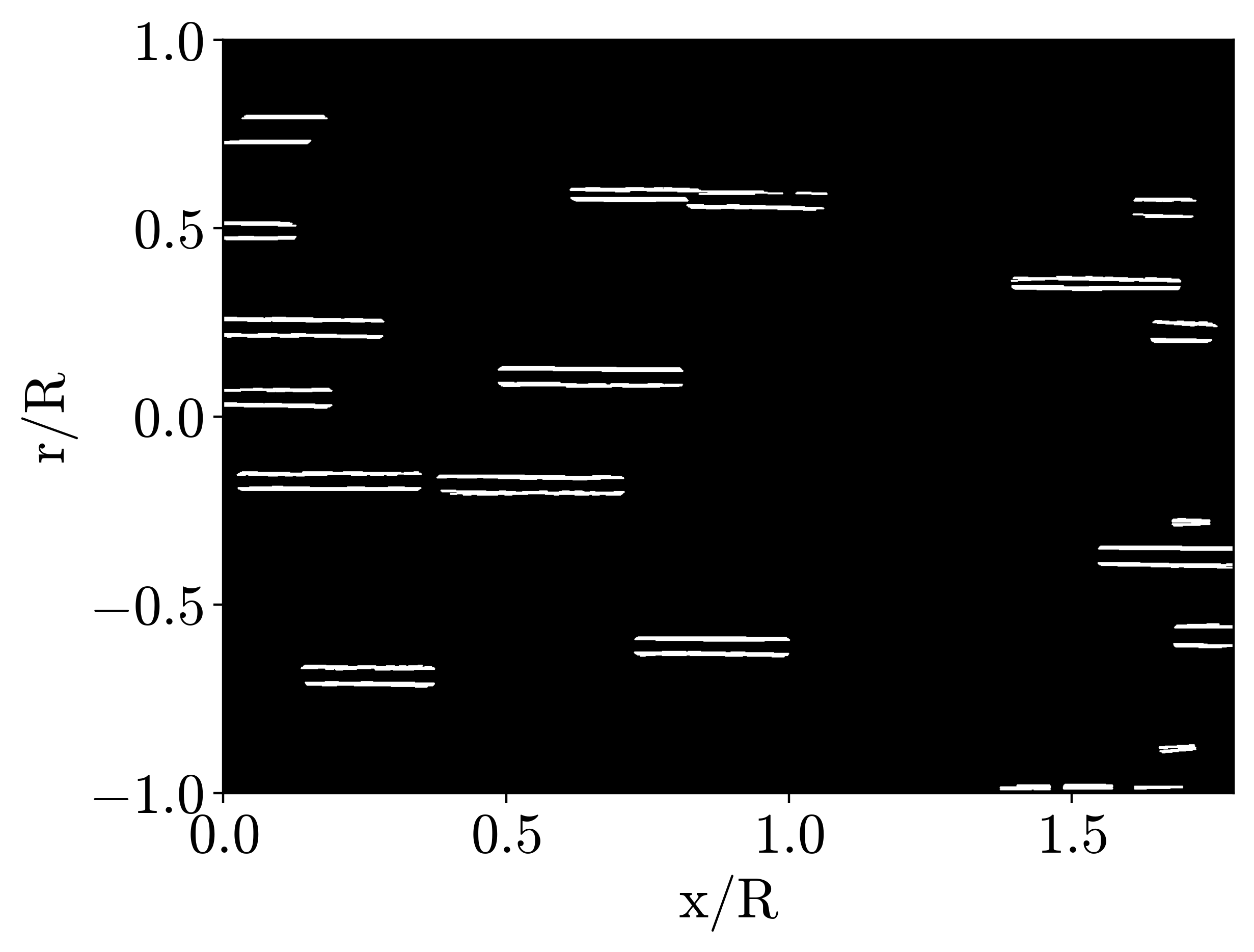}
        \caption{}
        \label{fig:hough_lines}
    \end{subfigure}
    \caption{Successive stages of image processing for the streak-angle velocimetry. In the vertical axis, r/R {means} a location r from the centreline for the given pipe radius R. In the horizontal axis, x/R {is} the distance x from the left of the image with respect to R. (a) Raw image, (b) background frame {obtained} from the average of all pictures within a run, (c) image after background subtraction, (d) image after adaptive thresholding, (e) image after canny edge detection and (f) Lines {detected} by Hough transform (showing the two lines counted as one per actual streak).}
    \label{fig:processing}
\end{figure}

\subsection{Experimental control parameters}
The problem {is governed} by the following non-dimensional numbers:
As in the non-particulate pipe flow, the Reynolds number $Re=UD/\nu$ {measures} the ratio of inertial to viscous forces outside the laminar regime and {is varied} in the range $[1120,2980]$.
The addition of neutrally buoyant spherical particles, of density $\rho$ and diameter $d_p$, {introduces} two additional parameters: the particle-to-fluid volume fraction (\(\Phi\)), {set} to \(\Phi=1.2 \times 10^{-3}\), and the Stokes number $St=\rho d_p^2 U/(18 \mu D)$ {measuring} the ratio of the particle’s relaxation time under the effect of drag exerted by the fluid, $\tau_p = \rho d_p^2/18\nu$, to the fluid’s advection time $\tau_U=D/U$. For the particles, $St$ {is} in the range $[0.2,0.6]$.
In practice, $Re$, $\Phi$, and $St$ {are controlled} through the piston's velocity that {imposes} the mass flux, the volume of particles introduced in the fluid, and the particles' diameter respectively.

Additionally, the flow perturbation {introduces} a radial mass flux through the perturbation inlet {controlled} dimensionally through the total volume of fluid introduced $V^\prime$, here {set to} $\SI{1}{ml}$ or $\SI{0.5}{ml}$, and the time of application of the perturbation $\tau^\prime$, here {set to} $\tau^\prime=\SI{90}{ms}$. To non-dimensionally measure the perturbation for a given $\tau^\prime$, we {introduce} the perturbation Reynolds number (\( \text{Re}_{\text{pert}} \)), which {is analogous} to the conventional Reynolds number and {is defined} using the averaged fluid velocity, obtained by dividing the volumetric perturbation flow rate with its pipe cross-sectional area calculated using its inner diameter. Two distinct values, $\text{Re}_{\text{pert}} = 3200$ and $\text{Re}_{\text{pert}} = 6400$, {corresponding} to perturbation volumes of 0.5 ml and 1 ml respectively, {are employed} in this study. \Cref{control_parameters} {provides} an overview of the control parameters {used} in this study.

\begin{table}[h]
    \centering
    \caption{Experimental Control Parameters}
    \label{control_parameters}
    \renewcommand{\arraystretch}{1.3} 
    \begin{tabular}{|p{5cm}|p{5cm}|}
        \hline
        \textbf{Parameter} & \textbf{Range / Value} \\ \hline
        Reynolds number (\( \text{Re} \)) & [1120, 2980] \\ \hline
        Particle-to-fluid volume fraction (\( \Phi \)) & \( 1.2 \times 10^{-3} \) \\ \hline
       Particle Stokes number (\( \text{St} \)) & [0.2, 0.6] \\ \hline
        Perturbation Reynolds number (\( \text{Re}_{\text{pert}} \)) & 3200, 6400 \\ \hline
    \end{tabular}
\end{table}

The parameters {are chosen} to cover a sufficiently large range of flow patterns for the purpose of their characterization by the streak-angle visualization method.

\subsection{Experimental procedure}
We {use} the following experimental procedure:
To initiate the experiment, the fluid-particles-tracers mixture {is circulated} upstream and downstream of the pipe to ensure particle and tracer homogeneity. The flow {is then gently reversed} to expel air bubbles, directing them out at the reservoir side. Afterward, the system {is left idle} for 25 minutes to allow the fluid to reach a standstill state.
The motor, controlling the piston-cylinder arrangement, {is then set} to a specific rotation speed to achieve the target Reynolds number $Re$.

Once the flow {is fully developed}---ensured by positioning the perturbation system 225 D downstream of the reservoir inlet \cite{durst2005development}---a controlled perturbation {is introduced}. This placement does not simply indicate that the flow has exited the acceleration phase; rather, it confirms that the flow has reached a steady, fully established state as dictated by the inlet conditions.

The first visualization system, combining Particle Tracking Velocimetry (PTV) and Particle Image Velocimetry (PIV), {captures} high-resolution flow images timed with the perturbation's arrival and {provides} us with a detailed analysis of the velocity field. Then, the second visualization system {is used} to perform streak visualization on flow patterns resulting from the downstream evolution of the flow within the fluid volume captured by the first system. Particle streaks {are recorded} to analyze downstream flow patterns and perturbation evolution.

The differential pressure transducer {continuously monitors} the pressure drop across the pipe during the entire run. The system {is reset}, the fluid {re-mixed}, and {allowed to settle} as described above between each run to ensure consistent experimental conditions.

\section{Image Processing for the Streak Visualization System}

The image processing for the streak visualization system {is carried out} using the OpenCV library in Python \citep{opencv-python}. The primary goal of this process {is} to detect the lines corresponding to the streaks visible in the image and then accurately {extract} the angles that these lines form with the pipe axis. This information {is crucial} for understanding the dynamics and flow characteristics within the pipe.

The initial step in this process {involves} capturing raw images from the camera during the experiment (Figure \ref{fig:raw_image}). These raw images often {contain} various artifacts caused by the reflection of light from the surface of the pipe, which {can interfere} with the accurate detection of streaks. To address this issue, a background frame (Figure \ref{fig:background_image}) {is constructed} by averaging all the frames captured during each experimental run. The purpose of this background frame {is to capture} the persistent elements of the image that are not of interest, such as static reflections and uniform lighting patterns. By subtracting this background frame from each raw image, we {obtain} a background-subtracted image (Figure \ref{fig:background_subtracted_image}), which {is significantly cleaner} and free from most artifacts. This subtraction step {is essential} to isolate the dynamic features, i.e., the streaks, from the static background.

After obtaining a cleaner, artifact-free image, the next step {is to identify} the streaks within the image. This {is accomplished} by converting the image into a binary format, where the pixels corresponding to the streaks {are assigned} a value of 1 (white), and the background pixels {are assigned} a value of 0 (black). This binarization {is typically achieved} through thresholding techniques. In simple or global thresholding, a single threshold value {is applied} across the entire image: pixels with values above this threshold {are set} to 1, and those below {are set} to 0. However, this approach {is often insufficient} for our application due to the uneven lighting and varying brightness of the streaks across different regions of the image.

To address these challenges, an adaptive thresholding algorithm {is employed}. Unlike global thresholding, adaptive thresholding {calculates} the threshold for each pixel based on the pixel values in its local neighborhood. In our methodology, we {use} a neighborhood defined by a square window with a side length of 101 pixels centered on the pixel of interest. The threshold for each pixel {is set} to the mean value of the pixels within this window minus a constant \(C\). For our experiments, we {have empirically determined} the optimal value of \(C\) to be 8. The images captured {have} an 8-bit depth, meaning each pixel {can have} an intensity value ranging from 0 (black) to 255 (white). The choice of the adaptive thresholding parameters, such as the neighborhood size and the constant \(C\), {is made} based on careful visualization and experimentation specific to our setup, as illustrated in Figure \ref{fig:adaptive_threshold}. These parameters {are critical} and {are influenced} by factors such as the bit depth, sensitivity, and resolution of the camera, as well as the illumination conditions during the experiment. In cases where the images {are noisy} or grainy, it {is advantageous} to apply Gaussian smoothing before performing adaptive thresholding. This smoothing step {reduces} noise and {enhances} the quality of the binary image, making the subsequent streak detection more accurate.

Once the binary image {is obtained}, the next step {involves detecting} the streaks as lines. A common approach for line detection {is} the Hough transform \citep{hough1962method, duda1972use}, which {is effective} in identifying lines in binary images. However, directly applying the Hough transform to our binary images {can lead} to multiple lines being detected for a single streak due to the non-negligible thickness of the streaks. This multiplicity {would skew} the analysis, as the number and angles of detected lines {would incorrectly reflect} the streaks' actual characteristics. To prevent this, we first {apply} the Canny edge detection algorithm \citep{canny1986computational}, which {is designed} to detect the edges of objects in an image. By detecting the edges of the streaks (Figure \ref{fig:edge_detection}), we {can then apply} the Hough transform to these edges, resulting in a more accurate representation of the streaks as lines. The Canny edge detection algorithm, as implemented in OpenCV, {involves} several parameters, including the minimum and maximum thresholds and the aperture size, which {determines} the size of the matrix used for the derivative computation. Since our image {is binary}, the results {are relatively insensitive} to the specific threshold values, and we {have selected} the smallest aperture size of 3 to ensure precise edge detection.

For the line detection itself, we {employ} a probabilistic variant of the Hough transform \citep{matas2000robust}. This variant {significantly reduces} the computational load by considering only a random subset of points in the image for line detection, with a minor trade-off in accuracy. The parameters for the OpenCV implementation of this probabilistic Hough transform {are carefully chosen} to balance detection accuracy and computational efficiency. Specifically, we {set} the distance resolution to 1 pixel, the angle resolution to 1 degree, an accumulator threshold of 30 votes, a minimum line length of 30 pixels, and a maximum line gap of 20 pixels. The term "accumulator threshold" {refers} to the minimum number of votes required for a line to be detected. Each vote {indicates} a point in the image supporting the presence of a line at a particular position and angle. Setting a threshold of 30 votes {ensures} that only prominent lines, with sufficient supporting points, {are identified}. These parameters {ensure} that the detected lines {correspond} closely to the actual streaks in the image while {minimizing} the detection of spurious lines due to noise or minor image imperfections.

In summary, the detection of streaks in images {involves} a sequence of well-defined steps: background subtraction to remove artifacts, adaptive thresholding to create a binary image, Canny edge detection to identify the edges of streaks, and the probabilistic Hough transform to detect the streaks as lines. Each step {is carefully tailored} to the specific needs of the experimental setup, including the characteristics of the imaging system and the environmental conditions. The empirical selection of parameters for adaptive thresholding, edge detection, and Hough transform {is crucial} for the robustness of the system, {ensuring} that the detected lines {accurately represent} the streaks and that the angles extracted {are reliable} for further analysis.

\section{Flow state identification based on the fluid phase}\label{sec3}
\subsection{Identification of flow pattern using PIV}
We {start} by characterizing the flow state using traditional techniques based on the full flow field delivered by PIV measurements from the first system. Patterns {are identified} by analyzing centerline velocity, velocity fluctuations, and average axial velocity profiles from PIV data. In the laminar phase, the average velocity profile {matches closely} with the Hagen-Poiseuille profile, indicating steady flow with minimal fluctuations. The transitional flow features such as puffs {are identified} by analyzing characteristic velocity fluctuations. Puffs {are detected} as localized turbulent structures with distinct leading and trailing edges (fronts and tails). As they {pass} a point along the pipe, they {cause} transient drops in centerline velocity while the flow upstream and downstream of a puff {remains} laminar. In the turbulent phase, the profile {is approximately aligned} to the 1/7th power-law profile, with high-frequency fluctuations and irregular spatial variations. These distinct profiles and centerline changes {enable} the precise identification of laminar, transitional, and turbulent flow states.

The flow patterns observed in response to \( \text{Re}_{\text{pert}} = 3200\) and \( \text{Re}_{\text{pert}} = 6400 \) ($\phi = 1.2 \times 10^{-3}$, particle size range = 212 -- 250 \si{\micro\metre}) when the fluid Reynolds number {is increased}, {are reported} in Table \ref{flow}.
For smaller $Re$, \emph{i.e.} $Re=1120$ and $Re=1530$, the flow {is} laminar. However, a transition to localized turbulence {takes place} at at most $Re=1980$, as puffs {are detected} for this value. In these cases, turbulent patches ('Puff') {appear} within an otherwise laminar flow, signifying the onset of turbulence within the system. For $Re\geq2260$, the turbulent nature of flow {takes over}.

The state of particulate pipe flows {is usually characterized} using the perturbation intensity 
$\epsilon$ = $(d_p/D)^{\frac{1}{2}}$ $(\phi)^{\frac{1}{6}}$. The critical Reynolds number \( \mathrm{Re}_{\mathrm{s,c}} \), marking the onset of turbulence, {scales} as \( \mathrm{Re}_{\mathrm{c}} \sim \epsilon^{-1} \). This inverse relationship {implies} that as particle perturbations grow (via increased particle size or concentration), turbulence {initiates} at lower Reynolds numbers \cite{hogendoorn2022onset}.
For the parameters we {consider}, $\epsilon\in[0.033, 0.0515]$, and so the transition {is expected} to take place in the range $Re_c\in[2200, 2400]$, when our measurements {capture} the presence of puffs at $Re = 1980$ and transition in the region $Re_c \approx 2260$, well within the expected range. On this basis, our results {are consistent} with the literature \cite{hogendoorn2022onset, hogendoorn2018particle, agrawal2019transition}.

\begin{table}[htbp]
    \centering
    \renewcommand{\arraystretch}{1.5}
    \medskip 
    \caption{Observed flow patterns in a particulate flow using PIV and Streak Visualization (described in \cref{streak}). 'Lam' and 'Turb' correspond to laminar and turbulent flow states extending over several diameters, while 'Puff' corresponds to a small turbulent patch, typically less than a pipe diameter long inside an otherwise laminar flow.}
    \label{flow}
    \begin{tabular}{|p{2.5cm}|*{6}{p{1cm}|}}
    \hline
    \multicolumn{7}{|c|}{Parameters: $\text{Re}_{\text{pert}} = 3200, 6400$, $\phi = 1.2 \times 10^{-3}$, Particle size range = 212--250 \si{\micro\metre}} \\
    \hline
    Reynolds Number & 1120 & 1530 & 1980 & 2260 & 2550 & 2980 \\
    \hline
    Flow State (PIV) & \textcolor{blue}{Lam} & \textcolor{blue}{Lam} & \textcolor{violet}{Puff} & \textcolor{red}{Turb} & \textcolor{red}{Turb} & \textcolor{red}{Turb} \\
    \hline
    Flow State (Streak Visualization) & \textcolor{blue}{Lam} & \textcolor{blue}{Lam} & \textcolor{violet}{Puff} & \textcolor{red}{Turb} & \textcolor{red}{Turb} & \textcolor{red}{Turb} \\
    \hline
    \end{tabular}
\end{table}

\subsection{Utilization of Pressure Drop Measurements to Characterize Particulate Pipe Flow States}
\label{sec:friction_coef}

Identifying flow patterns from measurements at a single observation point in the pipe {suffers} from uncertainty due to the evolution of the turbulence along the pipe: decaying regions of turbulence {cannot be distinguished} from expanding ones, so laminar and turbulent states {cannot be distinguished} with full certainty over long pipe ranges. The issue {is alleviated} by using the differential pressure meter to assess the pressure drop between the point where the perturbation {is injected} into the flow and the pipe's outlet. The pressure drop {is controlled} by the state of the fluid phase, and so {offers} a way to diagnose the turbulent or laminar state of the flow that {is independent} of the two visualization systems and crucially, regardless of how turbulent patches {may evolve} when traveling between the two systems. Of course, this method still {misses} the dynamics of turbulent patches decaying or growing over timescales much longer than the finite length of the rig {can capture}. 

Using the differential pressure measurements, we {calculate} the friction factor based on the formula provided in Eq. (\ref{1}):

\begin{equation}
f = \frac{2 D \Delta P}{\rho L U^2},
\label{1}
\end{equation}
where \(\Delta P\) {is} the measured pressure difference. 
We then {map} the experimentally determined friction factors with the theoretical friction factors for a pure fluid in both the laminar and turbulent regimes, as given by Eqs. (\ref{2}) and (\ref{3}), respectively.
In the laminar flow region, the friction factor {follows} the Hagen-Poiseuille law for a single fluid:
\begin{equation}
f_{\text{lam}} = \frac{64}{\text{Re}}
\label{2}
\end{equation}
In contrast, in the turbulent flow region, the friction factor for a smooth pipe {is described} by the Kármán–Prandtl resistance equation \cite{joseph2008friction}:
\begin{equation}
\frac{1}{\sqrt{f}} = 1.930 \log\left(\frac{\text{Re}}{\sqrt{f}}\right) - 0.537
\label{3}
\end{equation}
On this basis, we {obtain} friction from Eq. (\ref{1}) and {compare} it with those of (\ref{2}) and (\ref{3}) respectively. This {gives} an estimate of the flow state across the pipe length.

\begin{figure}[H]
    \centering
        \includegraphics[width=0.75\linewidth]{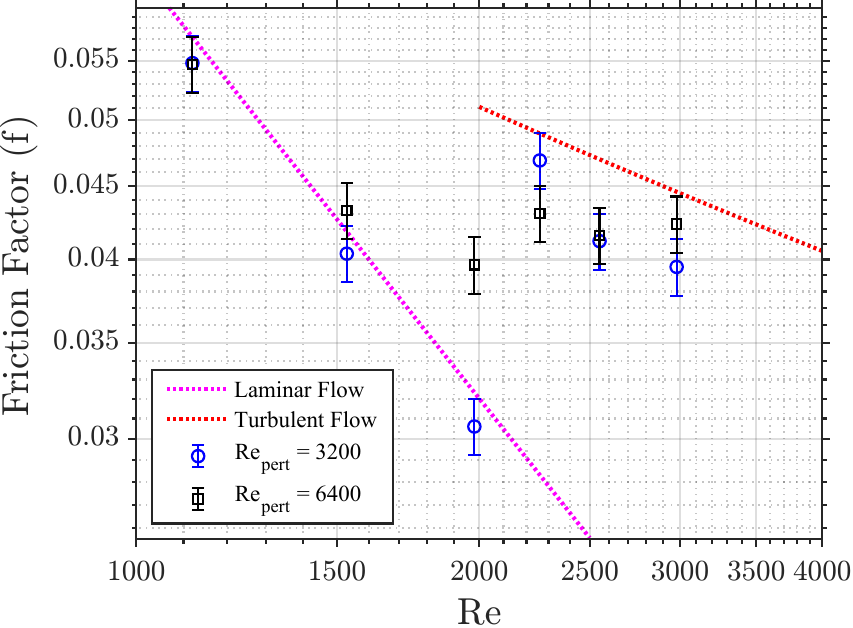}
    \caption{Friction factor vs $Re$ based on the pressure drop readings obtained from the particulate pipe flow experiment ($\phi = 1.2 \times 10^{-3}$, particle size range = 212--250 \si{\micro\metre}) against theoretical laminar and turbulent flows. }
    \label{friction_factor}
\end{figure}

The measured friction factors for the cases of  \( \text{Re}_{\text{pert}} = 3200\) and \( \text{Re}_{\text{pert}} = 6400 \) ($\phi = 1.2 \times 10^{-3}$, particle size range = 212 -- 250 \si{\micro\metre}) {are plotted} against $Re$ in Figure \ref{friction_factor}. For \( \text{Re}_{\text{pert}} = 3200\), $Re = 1120$ and $1530$, the friction {follows} the law for a single-phase laminar flow. At \(\text{Re} = 1980\), although the flow {exhibits} puffs detected by the Particle Image Velocimetry (PIV), see Section \ref{flow}, the friction still {coincides} with its predicted laminar value, meaning that the puff {was a localized feature} and {did not grow} over the pipe length. However, for Reynolds numbers of 2260 and higher, the friction {takes} a value between the laminar law and the friction law for a fully developed turbulent single-phase flow, indicating that the flow is turbulent in part of the region between the two pressure measurement points. When the friction is close to the turbulent friction value, the the flow between the two measurement points {is turbulent in most of the pipe}. 

Additionally, for \( \text{Re}_{\text{pert}} = 6400 \), the friction factor regime {is similar} to that obtained from \( \text{Re}_{\text{pert}} = 3200\) except for $Re = 1980$, indicating that for a higher $\text{Re}_{\text{pert}}$ the puff probably {survives and grows to fully developed turbulence}, and so affects the friction factor. 
Nonetheless, the intermediate value between the fully laminar and the fully turbulent one tells us that turbulence occupies only part of the pipe length, so the flow {is still} in the transitional flow regime and not fully turbulent. 

\begin{figure}[h!]
    \centering
    \begin{subfigure}[b]{0.64\textwidth}
        \includegraphics[width=\textwidth]{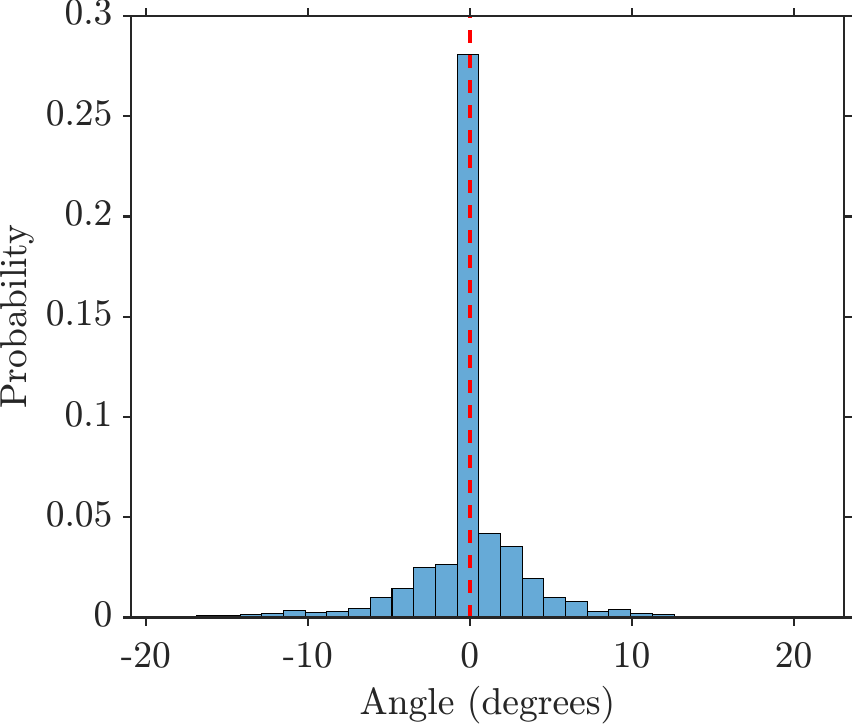}
        \caption{}
        \label{fig:img1}
    \end{subfigure}
    \hspace{0.5cm} 
    \begin{subfigure}[b]{0.64\textwidth}
        \includegraphics[width=\textwidth]{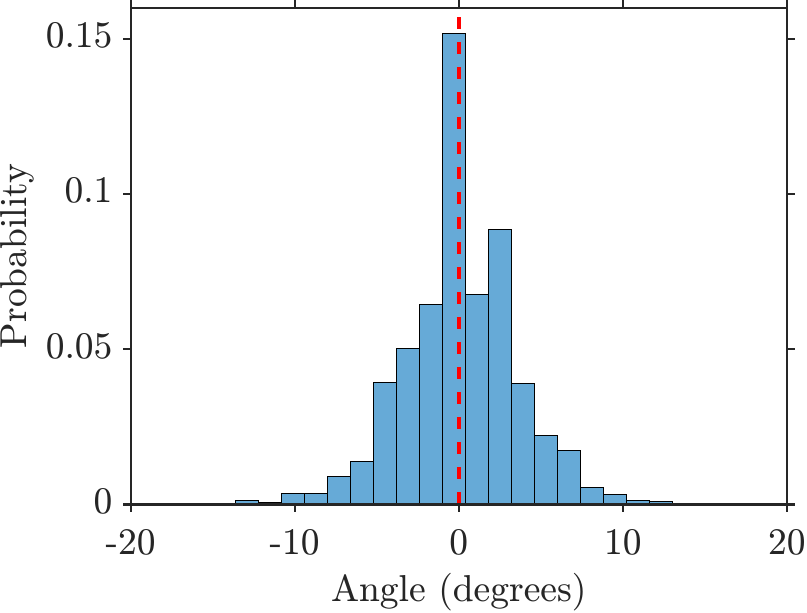}
        \caption{}
        \label{fig:img2}
    \end{subfigure}
    \caption{The depicted figures illustrate the probability distribution histograms of streak angles for two distinct Reynolds number scenarios ($\phi = 1.2 \times 10^{-3}$, particle size range = 212 -- 250 \si{\micro\metre}). Figure (a) {represents} the streak angle distribution for the case with $Re$ 1120, while Figure (b) {portrays} the streak angle distribution for the scenario with $Re$ 2980.}
    \label{angle_distribution}
\end{figure}

The results in this section {show} that combining PIV and pressure drop measurements both validate and complement each other. This {enables us to reliably diagnose} the flow state and {gives us confidence in using} them to validate the streak visualization method, which {is} the main point of this work.

\section{Characterisation of the flow states using streak-visualisation}
\label{streak}
\subsection{Streak visualisation}
A key difference between streak visualization and the techniques used in the previous section {is} that it {relies} on the movement of the particles, not the fluid phase. Hence, assessing the laminar or turbulent state of the flow in this manner {demands} that the motion of the particles {reflects} these states in an unequivocal way. 
While the particles (with $St\sim1$) do not follow the trajectories of the fluid, the main question is whether their trajectories display turbulent features (i.e. erratic, misaligned) when the fluid phase is turbulent and laminar (i.e. rectilinear and aligned in the streamwise direction when the fluid phase is laminar). 
Hence we first seek to characterise whether this is the case by intantaneous visualising snapshots of streaks corresponding to either flow regimes.

\begin{figure}[H]
    \centering
        \includegraphics[width=\linewidth]{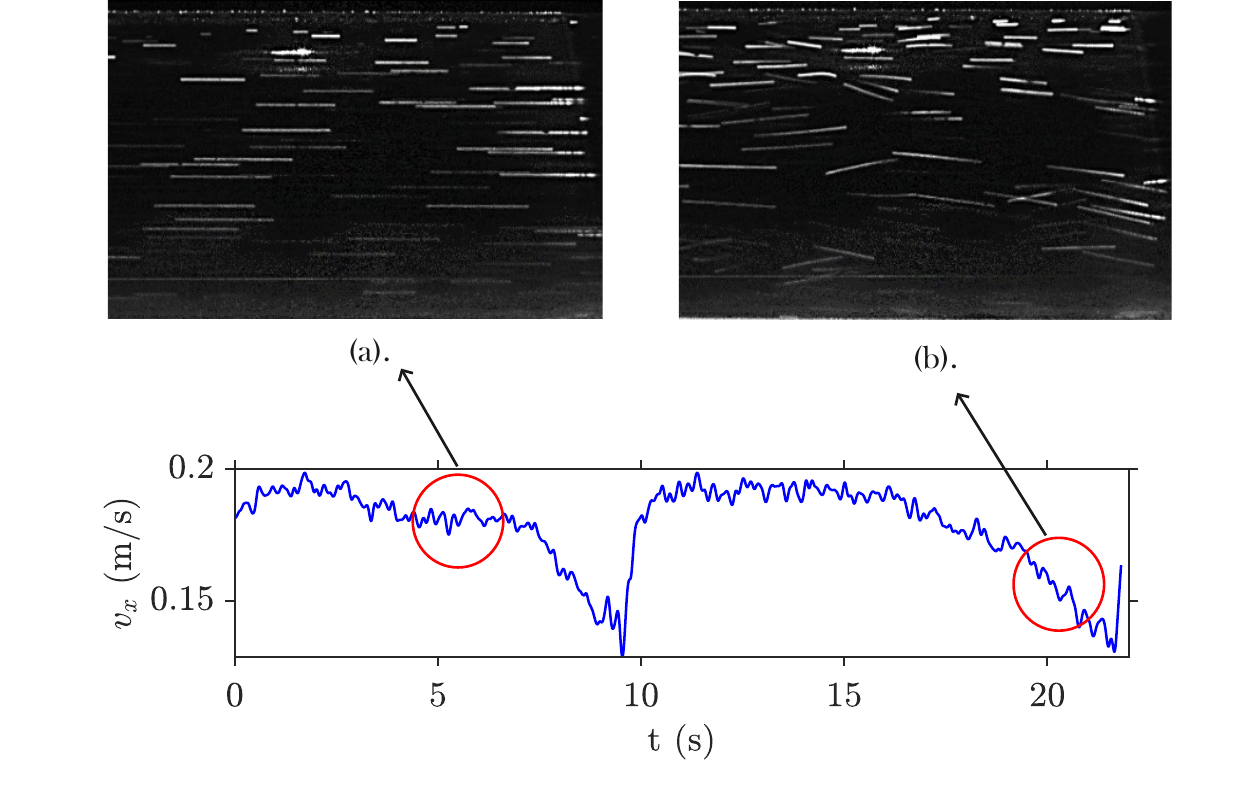}
    \caption{The blue line {is} the temporal centreline velocity of the fluid obtained from the PIV system at $Re$ 1980 ($\phi = 1.2 \times 10^{-3}$, particle size range = 212 -- 250 \si{\micro\metre}) showing the passing of two puffs. The two images (a and b) {shown here are obtained} directly from the camera of the streak visualization setup and {show} the streaks produced by the particles. Both images {are taken} at $Re$ 1980, but (a) {is} the part where the flow {is} laminar, and (b) {captures} the turbulent puff.}
    \label{puff}
\end{figure}

To this end, the images (a and b), as depicted in Figure \ref{puff}, {are captured} directly from the camera of the streak visualization setup, showcasing the streaks generated by the particles within the fluid flow. An example is shown in figure \ref{puff}. Two images {are shown} at $Re=1980$ ($\phi = 1.2 \times 10^{-3}$, particle size range = 212 -- 250 \si{\micro\metre}). Image \ref{puff}(a) {corresponds} to an instant where the flow is laminar, where streaks are qualitatively well aligned with the streamwise direction. Image \ref{puff}(b) {captures} one instant during the passage of a turbulent puff and shows streaks at finite angles with the streamwise direction. 
These images \ref{puff} (a and b) {are approximately taken} at positions indicated by red circles overlaid on the centerline velocity profile obtained from the PIV setup. This positioning {highlights} a visible variation in streak angles between the laminar and turbulent regions of the fluid flow. Specifically, the observed change in streak angle {serves} as a visual indicator of the dynamics over the short exposure time of the camera, which {compared} to the global flow timescale {is nearly instantaneous}. This different was noticeable in all examples and all instants where the flow could be clearly identified as being laminar or turbulent. As such, streak visualization based on the particles {potentially makes it possible} to follow the time evolution of the state of the fluid phase. To achieve this at every instant of a series of recorded images and monitor the evolution of the flow state in time, a large number of consecutive images {must be processed}, so we {need an quantitative criterion} to identify the flow state in each streak snapshot.

\subsection{Flow behavior identification based on the standard deviation of streak angles}
To quantify the distribution of streak angles with respect to the centreline, we {plot} the binned probability distribution of angles for two typical cases. The result, illustrated in \Cref{angle_distribution}, {demonstrates} a clear disparity in the angle distributions between cases for two different $Re$ numbers ($Re = 1120$ and $Re = 2980$ ($\phi = 1.2 \times 10^{-3}$, particle size range = 212 -- 250 \si{\micro\metre})). In the histogram representation, angles {cluster} towards zero degrees in laminar flows, signifying a more uniform directionality of particle streaks. Conversely, in relatively turbulent flows, the angle distribution {is notably wider}, reflecting the chaotic nature of particle motion within the fluid. This reflects the qualitative impression obtained by observing snapshots of streaks in laminar and turbulent flow states in the previous section.  Indeed, differentiating between the two cases {is often easy} by simply looking at the picture. There {are, however,} cases where transitional features {are difficult to differentiate}. Hence, there {is} a need for a more systematic approach to the detection of patterns from pictures. The simplest approach to this problem {is to imitate} "human recognition" in its most basic form: when assessing whether streaks {are mostly horizontally aligned} or more randomly distributed, we {assess} an average distribution and its scattering, \emph{i.e.}, the standard deviation of the angles (the average {is always expected} to be close to 0 provided there {is} a sufficiently large number of streaks in the picture considered).

However, to quantify the orientation of the streaks with sufficient statistical convergence, we {consider} the distribution of streak angles over multiple frames. The mean and standard deviation of these angles {are calculated} over a moving window of 5 frames. Increasing the number of frames used for computing the statistics {decreases} the error and noise in the estimates but {decreases} the temporal resolution of various features of interest. Decreasing the number of frames {has} the opposite effect. Hence, there {is} a trade-off between time resolution (which {is improved} by a higher sampling frequency, hence by reducing the number of successive frames used for the determination of the standard deviation) and statistical convergence.

We denote the set of streak angles obtained from 5 consecutive frames as $\{\theta_{i,i_n}\}$, where $i$ {varies} from 1 to $5$ {denotes} the frame and $i_n$ {denotes} the streak angles in the $i$\textsuperscript{th} frame. The standard deviation $\sigma$ of these values {measures} the dispersion or spread of the angles around their mean value. Mathematically, the sample standard deviation $\sigma$ {is calculated} as follows:
\begin{equation} \label{eu_eqn}
\sigma = \sqrt{\frac{\sum_{i=1}^{5}\sum_{i_n} (\theta_{i,i_n} - \bar{\theta}_i)^2}{\left(\sum_{i=1}^{5}\sum_{i_n}1\right)-1}},
\end{equation}
where $\bar{\theta}$ {is} the mean of the set of angles $\{\theta_{i,i_n}\}$ given by:
\begin{equation} \label{eu_eqn2}
\bar{\theta} =  \frac{\sum_{i=1}^{5}\sum_{i_n} \theta_{i,i_n}}{\sum_{i=1}^{5}\sum_{i_n}1}.
\end{equation}
A higher standard deviation $\sigma$ {indicates} greater variability in the streak angles across the frames. When the fluid flow {is} laminar, the streak angles {tend} to align parallel to the reference axis (e.g., horizontal), resulting in a lower standard deviation. In contrast, turbulent flow {leads} to irregular and chaotic movement of particles, causing streaks with varying angles, resulting in a higher standard deviation.

We {analyze} the angles of streaks produced by particles in a fluid, focusing on their distribution characteristics under laminar and turbulent flow conditions. The angles of these streaks {are indicative} of the underlying flow properties, with laminar flows typically {exhibiting} narrower angle distributions tending towards zero degrees, while turbulent flows {display} broader and more spread-out distributions.

\begin{figure}[H]
  \centering
  \begin{minipage}{0.46\textwidth}
    \centering
    \includegraphics[width=\textwidth]{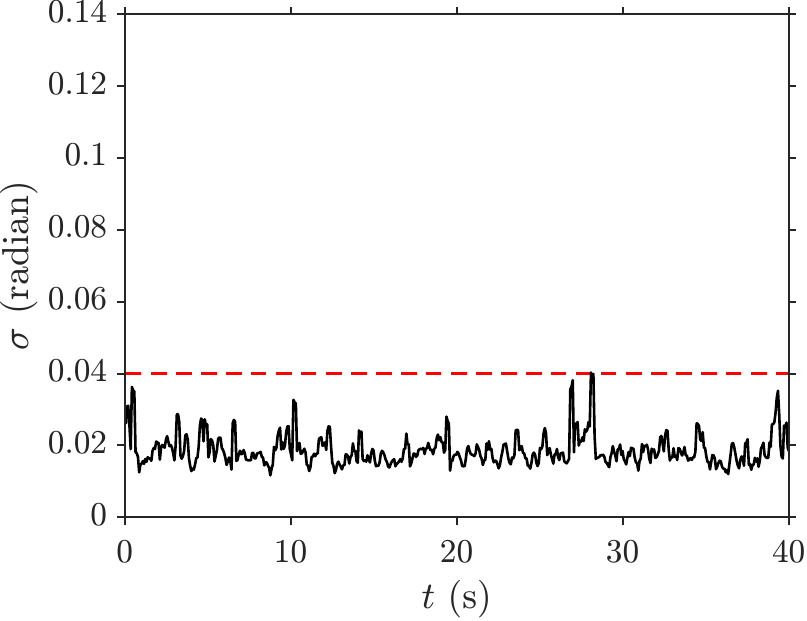}
    \subcaption{$Re$ = 1120}
  \end{minipage}
  \begin{minipage}{0.46\textwidth}
    \centering
    \includegraphics[width=\textwidth]{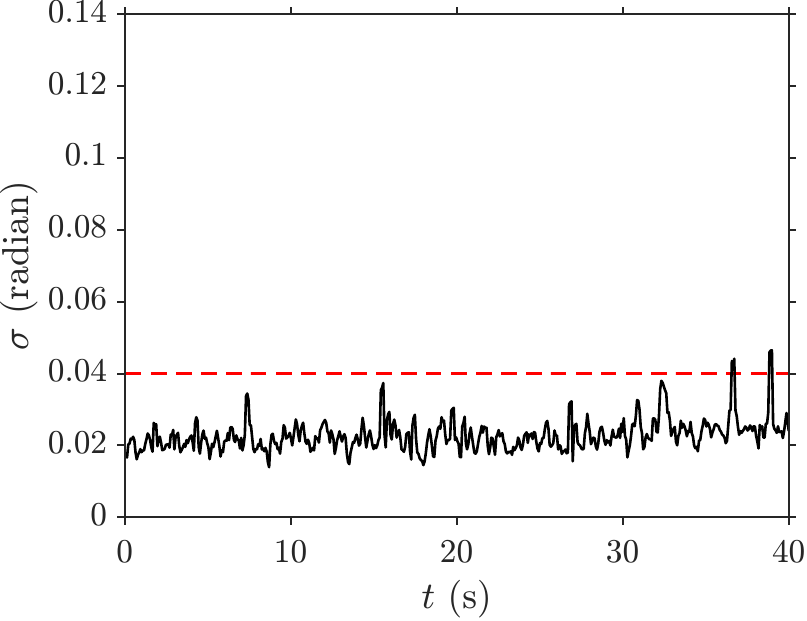}
    \subcaption{$Re$ = 1530}
  \end{minipage}

  \vspace{0.01cm}

  \begin{minipage}{0.46\textwidth}
    \centering
    \includegraphics[width=\textwidth]{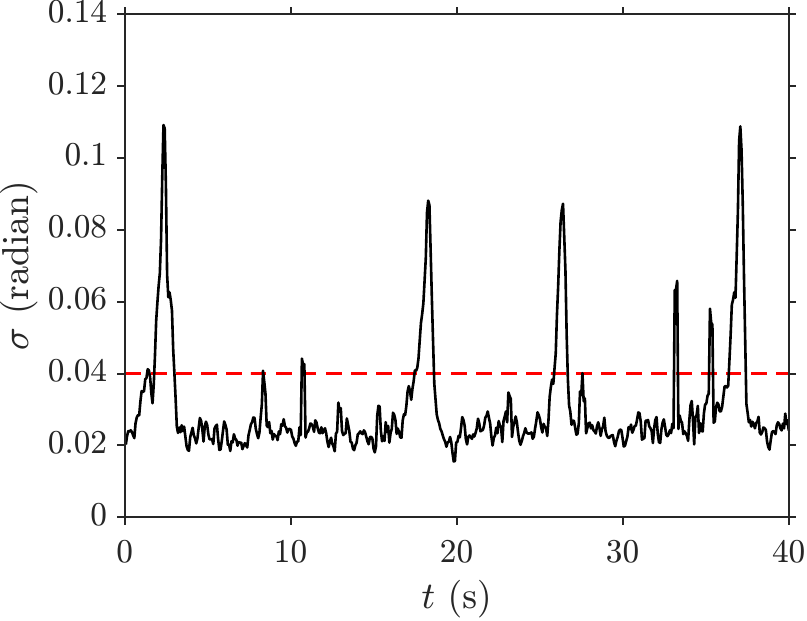}
    \subcaption{$Re$ = 1980}
  \end{minipage}
  \begin{minipage}{0.46\textwidth}
    \centering
    \includegraphics[width=\textwidth]{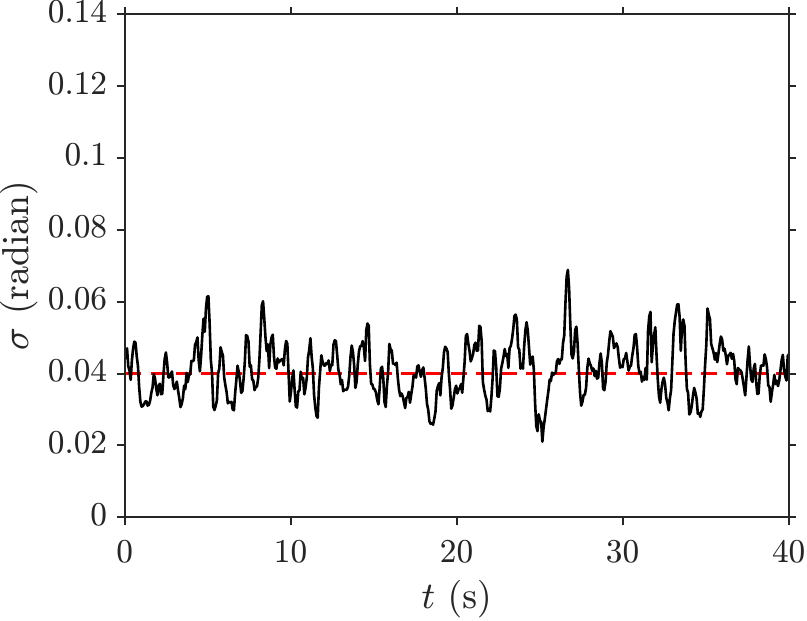}
    \subcaption{$Re$ = 2260}
  \end{minipage}

  \vspace{0.01cm}

  \begin{minipage}{0.46\textwidth}
    \centering
    \includegraphics[width=\textwidth]{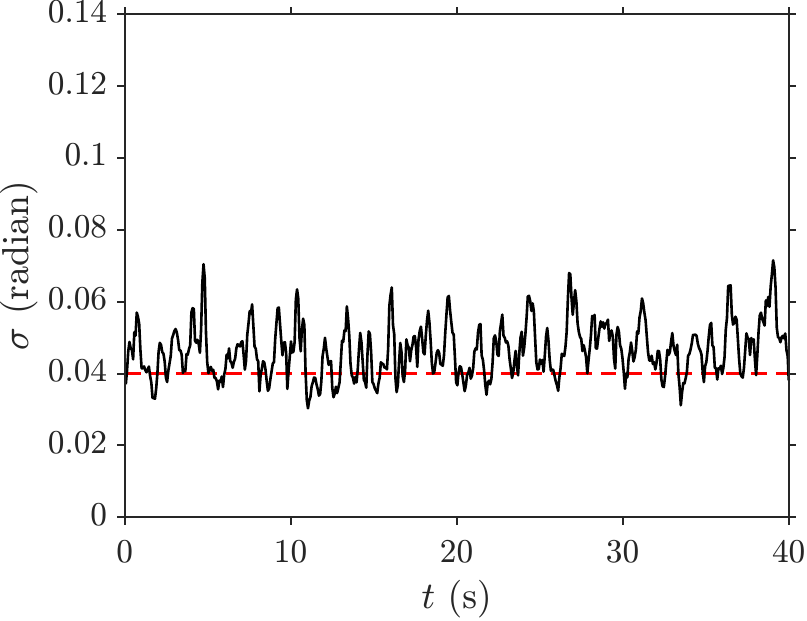}
    \subcaption{$Re$ = 2550}
  \end{minipage}
  \begin{minipage}{0.46\textwidth}
    \centering
    \includegraphics[width=\textwidth]{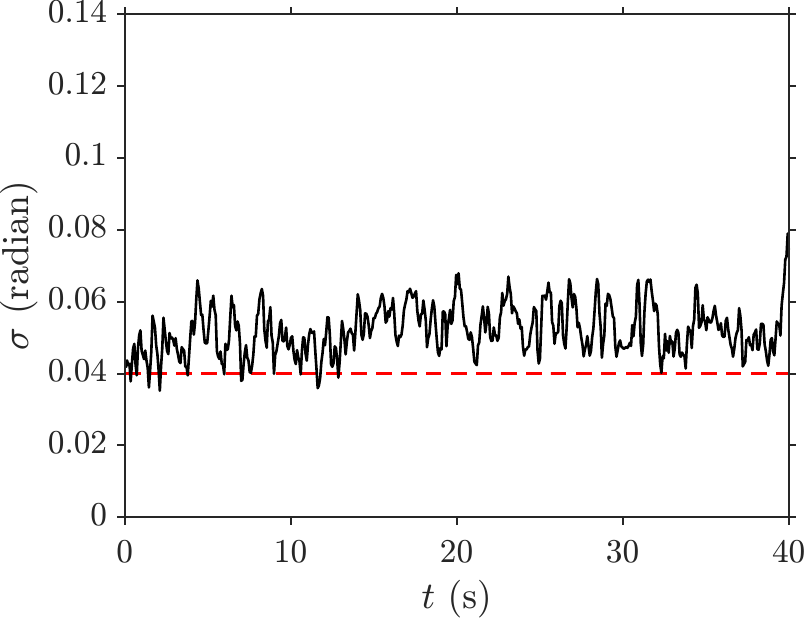}
    \subcaption{$Re$ = 2980}
  \end{minipage}

  \caption{Variation of Standard Deviation over Time for different Reynolds Numbers for fluid-particle mixture ($\phi = 1.2 \times 10^{-3}$, particle size range = 212 -- 250 \si{\micro\metre} ): Subfigures {depict} the temporal evolution of standard deviation values calculated from streak angles obtained through frame analysis for six different Reynolds numbers. The x-axis {represents} time (in seconds), while the y-axis {represents} the standard deviation (in radians) of streak angles. Additionally, a red line at the standard deviation value of 0.04 {serves} as a reference for distinguishing between laminar, transient, and turbulent flow regimes.}
  \label{results}
\end{figure}

We {first conduct} repeated measurements under conditions of relatively high and low Reynolds numbers ($Re=1120$ and $Re= 7500$), representing fully turbulent and fully laminar flows respectively to assess lower and upper bounds for the values of standard deviation. By plotting standard deviation values while simultaneously {validating} them with the PIV setup, a reference red line ($\sigma = 0.04$) {separating} both cases {is established} (as depicted in \Cref{results}), {serving} as a benchmark for less obvious states between the fully turbulent and fully laminar reference cases. We also {cross-check} the nature of the flow for each value of $Re$ presented in \Cref{results} using the PIV system and the pressure drop system to further {confirm} the effectiveness of using the statistical distribution of streak angles in precisely understanding the flow characteristics.

The examination of subfigures in Figure \ref{results} {reveals} more detailed information on the flow dynamics: For Figure \ref{results}(a) and (b), the flow {is} laminar (the standard deviation value throughout the time {is} well below the red line demarcation). As for the case within the transitional regime, found around $Re=1980$, shown in Figure \ref{results}(c), clear peaks of standard deviation {indicate} intermittent turbulent patches (puffs) amidst predominantly laminar flow conditions: these {are also observed} through the PIV system. As the flow gradually transitions towards turbulence, exemplified by subfigure \Cref{results}(d), the standard deviation progressively {approaches} and eventually {surpasses} the red line reference, {indicating} the onset of turbulent behavior. The turbulence {invades} a greater part of the signal and finally all of it in figures \Cref{results}(e) and \Cref{results}(f), where the standard deviation consistently {exceeds} the red line, {signifying} fully turbulent flow conditions.

The outcomes of the flow analysis conducted with the pressure drop setup (\cref{friction_factor}) and the PIV system (\cref{flow}) {closely match} the conclusions given by the proposed visualization system. The analysis and comparison of each case to the reference plot shown in Figure \ref{results} {reinforce} the identified trends in flow dynamics by the streak visualization method in not only identifying the nature of the flow but also detecting the presence of any transitional flow feature. Here, the identification of a simple threshold in the value of the standard deviation, calibrated from the purely laminar and purely turbulent state, {makes it possible} to reliably detect transient features such as puffs.

%
\subsection{A new approach to the estimation of critical Reynolds number}
The probability distribution of angles {contains} far more information on the flow than we {have used} so far. They {contain} a measure of how far the flow {stands} from the laminar or the fully developed turbulent state. Until now, we {have only used} a threshold on the standard deviation to detect whether any turbulence {was present} but without quantifying "how turbulent" the flow {was}. Doing this {enables us to find} a threshold where the flow {becomes "closer"} to the turbulent state than to the laminar state in the spirit of the approach proposed by \cite{avila2011onset}\cite{mukund2018critical} to define a critical Reynolds number for the transition to turbulence. To quantify the idea of proximity between states, we {propose utilizing} the Kullback-Leibler (K-L) divergence. This statistical measure {provides} a means to quantify the difference between two probability distributions, offering a rigorous method for comparing observed angle distributions with reference distributions representing pure laminar and pure turbulent cases.

The KL divergence between two continuous distributions $P$ and $Q$ {is given} by:
\begin{equation}
    D_{\mathrm{KL}}(P \| Q)=\int_{-\infty}^{\infty} p(x) \log \left(\frac{p(x)}{q(x)}\right) \mathrm{d} x.
\end{equation}
Here \( p(x) \) and \( q(x) \) {represent} the probabilities of outcome \( x \) under distributions \( P \) and \( Q \). By employing K-L divergence, we {can effectively evaluate} the degree of similarity or dissimilarity between a given angle distribution and the reference distributions \cite{basseville2013divergence}. This approach {enables} a robust assessment of flow conditions based on the observed angle data, {facilitating} the identification of laminar, turbulent, or intermediate regimes within the fluid.

\begin{figure}[h]
    \centering
    \includegraphics[width=0.70\textwidth]{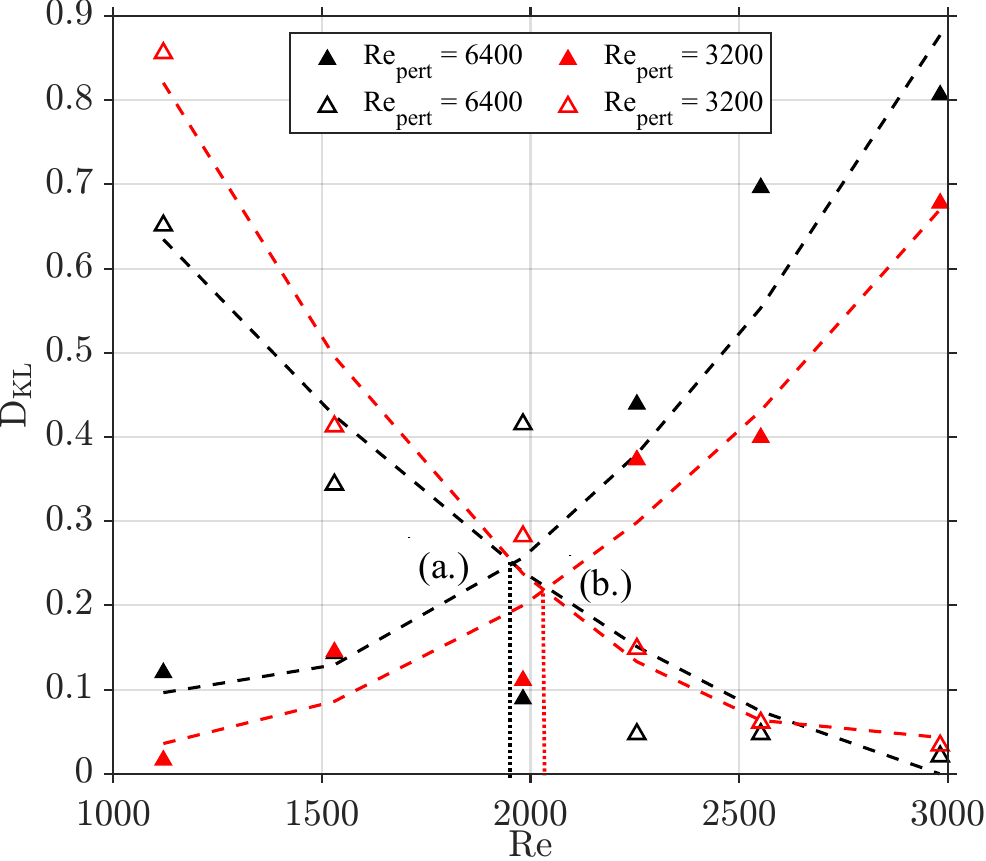} 
    \caption{K-L divergence for the above-mentioned $Re_{\text{pert}}$ having $\phi = 1.2 \times 10^{-3}$ and particle size range = 212 -- 250 \si{\micro\metre} where working fluid {is introduced} as perturbation upstream of the setup. The solid markers in the illustration {represent} instances where a laminar flow streak angle probability distribution with $Re=1120$ {is taken} as the reference case for calculating the K-L divergence in subsequent scenarios. Conversely, the hollow markers {indicate} situations where an $Re$ of 4500 {is adopted} as the reference case.}
    \label{k_l_div}
\end{figure}

As shown in \Cref{k_l_div}, for both the case of \( \text{Re}_{\text{pert}} = 3200 \) and \( \text{Re}_{\text{pert}} = 6400 \), the values of KL divergence ($D_{KL}(P|Q)$) {increase} with increasing $Re$ when the reference probability distribution $Q$ {is taken} from the laminar flow ($Re=1120$). $D_{KL}(P|Q)$ {takes} higher values for higher $Re$ as the presence of turbulence {distorts} the flow state away from the reference laminar state. At smaller $Re$, where flow {is mostly laminar}, the streak angle distribution {is similar} to that of the reference case at $Re=1120$.

Conversely, the K-L divergence values {decrease} with increasing $Re$ when the reference distribution $Q$ {is} that of a turbulent case $Re=4500$. Laminar states at low $Re$ {return} a high K-L divergence, and those at higher $Re$ {remain close} to 0, indicating a close match with the streak angle distribution of the $Re=4500$, which {is turbulent}.
 
Hence, the K-L divergence {provides} a quantitative measure of the turbulent state based on the knowledge of particle angles of a reference case of either a pure laminar or pure turbulent flow scenario.

To estimate the critical Reynolds number $Re_c$ using the graph (\Cref{k_l_div}), we {sought} a location where the K-L divergence {exhibits} a pronounced shift for both choices of laminar and turbulent reference distributions \( Q \).

Specifically, the graph {shows} that for both the laminar ($Re=1120$) and turbulent ($Re=4500$) reference cases, there {is} a distinct crossover region where the divergence values {change rapidly}, right around the point of intersection of K-L values calculated with either reference for each case. For the case of \( \text{Re}_{\text{pert}} = 3200\), this crossover {occurs} at a higher $Re$ (see \cref{k_l_div} (b) in the graph), \textit{i.e.}, \(Re> 2000\), {suggesting} that the transition to turbulence {happens} at a higher Reynolds number with smaller \( \text{Re}_{\text{pert}} \). This {is confirmed} by the friction factor measurements shown in \ref{friction_factor}, where the friction factor {corresponds} to that in the turbulent regime for \( \text{Re} > 2000 \). Conversely, for the higher perturbation case of \( \text{Re}_{\text{pert}} = 6400 \), the crossover {occurs} at a lower $Re$ (see \cref{k_l_div} (a) in the graph), \textit{i.e.}, \(Re< 2000\). This {indicates} that larger upstream perturbations {induce} an earlier transition to turbulence, which {is also confirmed} by the \ref{friction_factor} graph showing the friction factor in the transitional regime at these Reynolds numbers. This observation is consistent with the experimental findings showing the Friction factor vs $Re$ graph by Ref. \cite{hogendoorn2022onset}.

Hence, using the laminar and turbulent states as references offers a robust method to assess the point of transition to turbulence and the corresponding critical Reynolds number, using the K-L divergence. 
\section{Conclusion}

The study demonstrates that the streak visualization technique reliably distinguishes between laminar, transitional, and turbulent flow regimes in particulate pipe flows through the quantification of streak angles, supported by validation from PIV and pressure drop measurements.

This finding underscores the potential for accessible and cost-effective methods to complement or replace traditional, resource-intensive techniques, broadening their field of application both in research and to practical applications in fluid dynamics

The results demonstrate that the streak visualization technique is an effective tool for distinguishing between flow regimes in particulate pipe flows. By analyzing streak angle distributions, the method reliably identifies laminar, transitional, and turbulent states. Validation through complementary methods, including Particle Image Velocimetry (PIV) and pressure drop analysis, further supports its accuracy. The use of statistical measures, such as the standard deviation of streak angles and the Kullback-Leibler divergence, enhances the ability to detect flow transitions and determine the critical Reynolds number, effectively addressing the challenge of distinguishing erratic trajectories resulting from underlying turbulent fluid motion from regular trajectories associated to a laminar flow.

Some limitations remain, 
in particular the need for manual steps in image processing, which introduce subjectivity. However, automation of these steps and refinements to improve the system’s robustness are not expected to pose a major challenge and could extend the application of this method to a wider range of scenarios.
As such future developments of the method itself could focus on automated streak angle analysis and exploring applications in non-circular geometries or single-phase flows to expand the method’s utility. 

The findings align with existing literature on turbulence and transitional flow dynamics, in highlighting the role of particle behavior in influencing flow states. Compared to traditional methods like PIV and Laser Doppler Velocimetry (LDV), the streak visualization technique offers a more accessible and cost-effective alternative for the purpose of detecting laminar and turbulent flow patterns, without sacrificing reliability. This broadens its potential application in cases where resources for high-cost methods are unavailable.

However, the main scope to expand this method probably lies in the processing of the data it generates: while using a threshold on the standard deviation obtained by calibration against fully laminar and fully turbulent states offers a reliable, objective method to identify flow patterns, more advanced statistical analysis could be used to eliminate the calibration step. Indeed, despite its simplicity, streak velocimetry still collects extensive time-dependent flow data that was not used for the purpose of detecting flow patterns. The reliable definition of a transitional Reynolds number based on the KL divergence, certainly offers a glimpse of how more refined statistical analysis of the time-dependent statistical distribution of streak angles may provide deeper insight into the flow dynamics. 


\section*{Acknowledgements}
The authors would like to extend their gratitude to Ian Bates, lab technician at Coventry University, for his invaluable assistance. Special thanks to Lyse Brichet, intern (2022)  and Mathilde Schneider, intern (2020–2021) from Ecole Normale Supérieure de Lyon, Université,  for their assistance with data collection and ideation respectively. Additionally, the authors acknowledge the fluids research group of Fluids and Complex Systems at Coventry University for their support and collaboration.


\end{document}